\documentclass[lettersize,journal]{IEEEtran}
\usepackage{amsmath,amsfonts}
\usepackage{algorithmic}
\usepackage{algorithm}
\usepackage{array}
\usepackage[caption=false,font=normalsize,labelfont=sf,textfont=sf]{subfig}
\usepackage{textcomp}
\usepackage{stfloats}
\usepackage{url}
\usepackage{verbatim}
\usepackage{graphicx}
\usepackage{cite}
\begin{document}

\title{A Novel $\alpha\beta$-Approximation Method Based on Numerical Integration for Discretizing Continuous Systems}

\author{Shen Chen,~\IEEEmembership{Student Member,~IEEE,} Chaohou Liu, Wei Yao, Jisong Wang, Shuaipo Guo, Zeng Liu,~\IEEEmembership{Senior Member,~IEEE,} Jinjun Liu,~\IEEEmembership{Fellow,~IEEE}
\thanks{This paper was produced using the \LaTeX template released by the IEEE Publication Technology Group.}
\thanks{ Manuscript received January 30, 2026; 
This work was supported by the National Key Research and Development Program of China under Grant 2023YFB2604600.
(Corresponding author: Jinjun Liu.)

Shen Chen is with the School of Electrical Engineering, Xi'an Jiaotong University, Xi'an 710049, China,
and also with SolaX Power Network Technology (Zhejiang) Co., Ltd., Hangzhou 310013, China 
(e-mail: chenshen@stu.xjtu.edu.cn).

Zeng Liu, and Jinjun Liu are with the State Key Laboratory of Electrical Insulation and
Power Equipment, School of Electrical Engineering, Xi'an Jiaotong University, Xi'an 710049, China 
(e-mail: zengliu@mail.xjtu.edu.cn; jjliu@mail.xjtu.edu.cn).

Chaohou Liu, Jisong Wang, and Shuaipo Guo are with
SolaX Power Network Technology (Zhejiang) Co., Ltd., Hangzhou 310013, China 
(e-mail: liuchaohou@solaxpower.com; wangjisong@solaxpower.com; guoshuaipo@solaxpower.com).

Wei Yao is with the Key Laboratory of Technology in Rural Water Management of Zhejiang Province, 
College of Electric Engineering, Zhejiang University of Water Resources and Electric Power, Hangzhou 310018, China
(e-mail: yaowei@zjweu.edu.cn).
}
}

\markboth{IEEEtran \LaTeX\ Template v1.8b,~Vol.~x, No.~x, October~2025}%
{Shell \MakeLowercase{\textit{et al.}}: A Sample Article Using IEEEtran.cls for IEEE Journals}


\maketitle

\begin{abstract}
In this article, we propose a novel discretization method based on numerical integration for discretizing continuous systems,
termed the $\alpha\beta$-approximation or Scalable Bilinear Transformation (SBT).
In contrast to existing methods,
the proposed method consists of two factors, i.e., shape factor ($\alpha$) and time factor ($\beta$).
Depending on the discretization technique applied, 
we identify two primary distortion modes in discrete resonant controllers: frequency warping and resonance damping.
We further provide a theoretical explanation for these distortion modes, 
and demonstrate that the performance of the method is superior to all typical methods.
The proposed method is implemented to discretize a quasi-resonant (QR) controller on a control board, 
achieving 25\% reduction in the root-mean-square error (RMSE) compared to the SOTA method. 
Finally, the approach is extended to discretizing a resonant controller of a grid-tied inverter. 
The efficacy of the proposed method is conclusively validated through favorable comparisons among 
the theory, simulation, and experiments.
\end{abstract}

\begin{IEEEkeywords}
Discretization, $\alpha\beta$-approximation, numerical integration, scalable hexagonal approximation, shape factor, time factor, resonant controller.
\end{IEEEkeywords}

\section{Introduction}
\IEEEPARstart{D}{igital} control technology is a revolutionary force in modern industry, 
enabling high levels of automation. 
The transition from analog to digitally-controlled systems for superior accuracy and flexibility 
relies on a critical step: discretization. 
This process is not merely a technicality but a central design choice with profound implications.
In general, discretization methods can be categorized into two broad paradigms: 
the direct discrete design \cite{Direct-01}--\cite{Direct-03}, 
applied to systems that are inherently discrete, 
and the indirect discrete design \cite{Euler-01}--\cite{Exact-02}, 
in which a controller is first designed in the continuous-time domain ($s$-domain) 
and subsequently transformed into the discrete-time domain ($z$-domain) via an $s$-to-$z$ transformation. 
It is important to emphasize that all discretization methods introduce inherent errors, 
such as magnitude inaccuracies and phase distortions. 
More critically, some methods can compromise the stability of the resulting discrete system. 
Therefore, the strategic selection of an appropriate discretization method plays a pivotal role 
in determining the performance, reliability, and stability of the final digital implementation.

The resonant controllers are designed to achieve zero steady-state error 
at specific frequencies by leveraging a theoretically infinite gain at the system's resonant frequency. 
The core principle lies in the internal model principle,
which states that including a model of the reference or disturbance in the controller structure enables perfect tracking or rejection.
Due to their exceptional ability to track or reject sinusoidal signals with high fidelity, 
these controllers are widely employed in low-voltage dc (LVdc) distribution networks \cite{RC-01},
active power filters (APFs) \cite{RC-02}, \cite{RC-04}, \cite{RC-08}, 
voltage-controlled voltage-source inverters (VSIs) \cite{RC-03},
permanent magnet synchronous motors (PMSMs) \cite{RC-05}, \cite{RC-06}, \cite{RC-09},
and grid-connected inverter \cite{RC-10}, etc.

\begin{figure}[h]  
  \centering
  \includegraphics[width=1.0\linewidth]{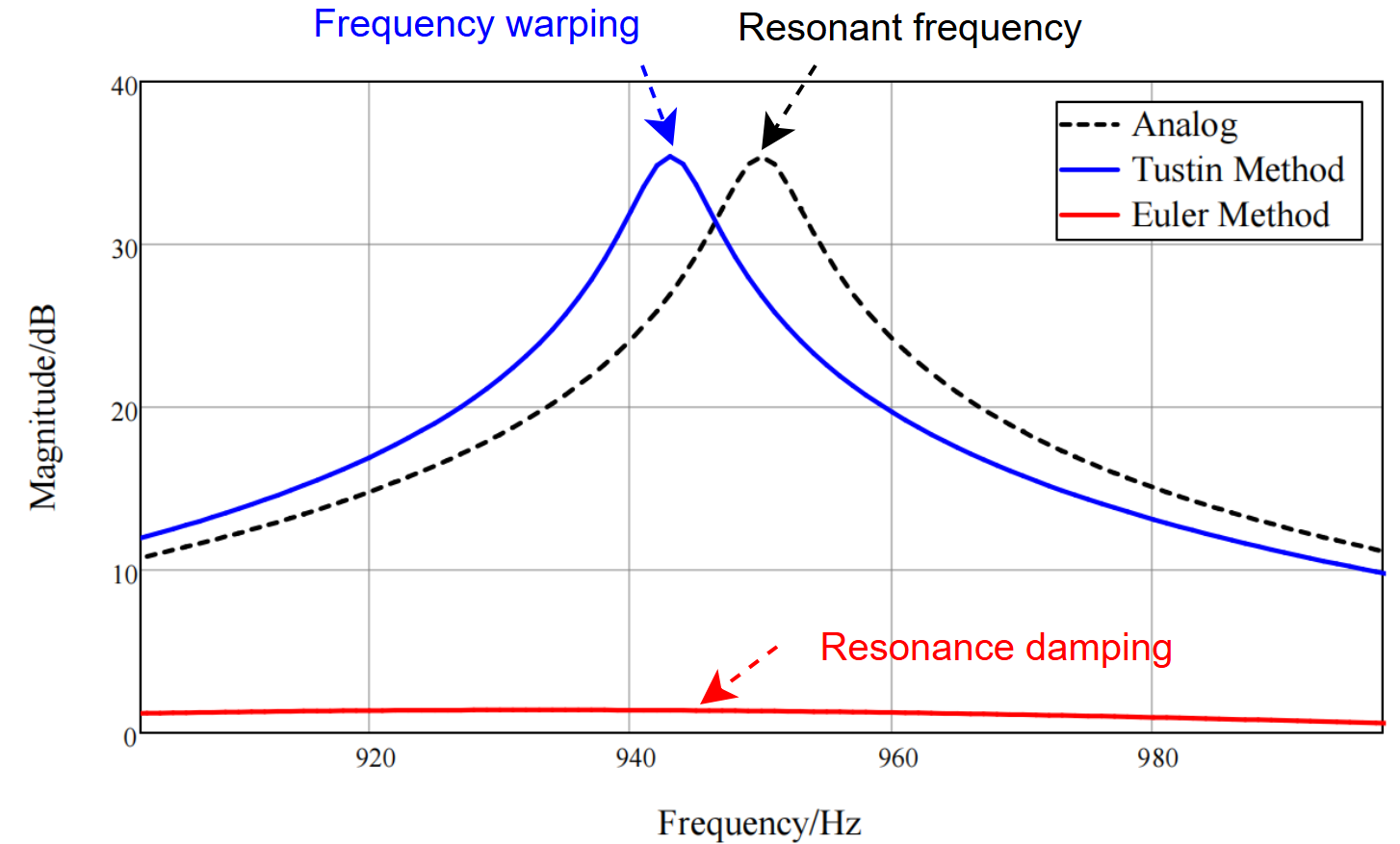}
  \caption{Bode plot of quasi-resonant controller under different discretization methods. (resonant frequency = 950 Hz, sampling frequency =20 kHz)}
  \label{fig_freq_warping}
\end{figure}

However, the performance of digital resonant controllers is highly sensitive to the choice of discretization method. 
Even minor deviations in the resonant pole location introduced during discretization can lead to significant performance degradation, 
which becomes increasingly pronounced as the resonant frequency approaches the Nyquist frequency. 
For instance, when applying the Euler or Tustin method 
(two widely adopted indirect discretization approaches favored for their simplicity and practical utility, 
accounting for approximately 64.2\% and 9.9\% of applications as summarized in \cite{Disc-Review}, respectively),
the Euler method exhibits pronounced resonance damping (red solid line in Fig.~\ref{fig_freq_warping}), 
while the Tustin method introduces a frequency-warping artifact near the resonant frequency (blue solid line). 
As illustrated in Fig.~\ref{fig_freq_warping}, at a resonant frequency of 950 Hz, 
the Tustin method results in a frequency shift exceeding 3 Hz, 
whereas the Euler method causes a magnitude attenuation of over 34 dB, 
ultimately leading to steady-state error. 
To mitigate these issues, various strategies such as pole correction and delay compensation have been proposed  \cite{RC-07} -- \cite{RC-09}, 
and a comprehensive comparison of discretization techniques for resonant controllers is available in \cite{RC-04}.
Among these methods, the Tustin method with pre-warping, referenced in \cite{RC-02}, \cite{RC-03}, and \cite{RC-05} 
has emerged as the state-of-the-art (SOTA) approach for implementing digital resonant controllers. 
For instance, \cite{RC-02} conducted an analysis of discretization methods for active power filters, 
and proposed a computationally efficient realization utilizing Tustin discretization with pre-warping compensation. 
Similarly, \cite{RC-03} compared three methods—Tustin with pre-warping, a two-integrator-based scheme, 
and the zero-order hold (ZOH) method—for voltage-controlled voltage-source inverters. 
Their results demonstrated that the Tustin method with pre-warping introduced no additional phase lag, 
whereas the other two methods incurred an extra time delay. 
Furthermore, \cite{RC-05} validated the effectiveness of the Tustin method with pre-warping through frequency-domain analysis. 
In a related effort, \cite{RC-07} proposed a modified Tustin method aimed at eliminating frequency deviation 
by replacing the resonant poles generated via the standard Tustin method with accurate resonant poles. 
Although this method is conceptually similar to the SOTA approach, it increases algorithmic complexity.

To the best of author's knowledge, the existing literature has not yet conclusively elucidated 
the fundamental reasons behind the performance variations observed 
when different discretization methods are applied to resonant controllers. 
Although \cite{RC-10} identified the pre-warped Tustin method as the most suitable choice for proportional-resonant (PR) controllers, 
the underlying factors responsible for its superiority remain unclear. 
Furthermore, it is worth exploring whether there is potential to surpass current SOTA techniques. 
Motivated by these open questions, 
this study seeks to provide a theoretical explanation for the performance differences 
and to develop an enhanced bilinear transformation method that exceeds the performance of existing SOTA approach.
The proposed method, termed $\alpha\beta$-approximation or Scalable Bilinear Transformation (SBT), 
is inspired by the Generalized Bilinear Transformation (GBT) introduced by Sekera in 2005 as the $\alpha$-approximation \cite{GBT-01-Sekara}. 
Unlike the GBT \cite{GBT-01-Sekara} -- \cite{GBT-03}, 
the SBT incorporates an additional scaling factor ($\beta$) to adjust the effective sampling time. 
This modification allows more precise tuning of the equivalent poles of the resonant controller, 
thereby improving accuracy and control performance.

The structure of this article is organized as follows. 
The paper begins by introducing a novel bilinear transformation method and elucidating its relationship with existing discretization approaches. 
Subsequently, the proposed SBT is applied to the discretization of a quasi-resonant (QR) controller. 
The discretization error, particularly in terms of magnitude distortion, 
is then explored through an analysis of pole variations under different methods, including Euler, Tustin, SOTA, and the proposed SBT. 
Following the theoretical analysis, the SBT is implemented to discretize a QR controller on a control board. 
Finally, the method is extended to the discretization of a PI+QR controller within a grid-tied inverter system. 
The efficacy of the proposed SBT is validated by comparing theoretical predictions with both simulation and experimental results.

\section{The Scalable Bilinear Transformation}
\subsection{Derivation of the SBT Based on Numerical Integration}
\begin{figure}[htbp]  
  \centering
  \includegraphics[width=0.8\linewidth]{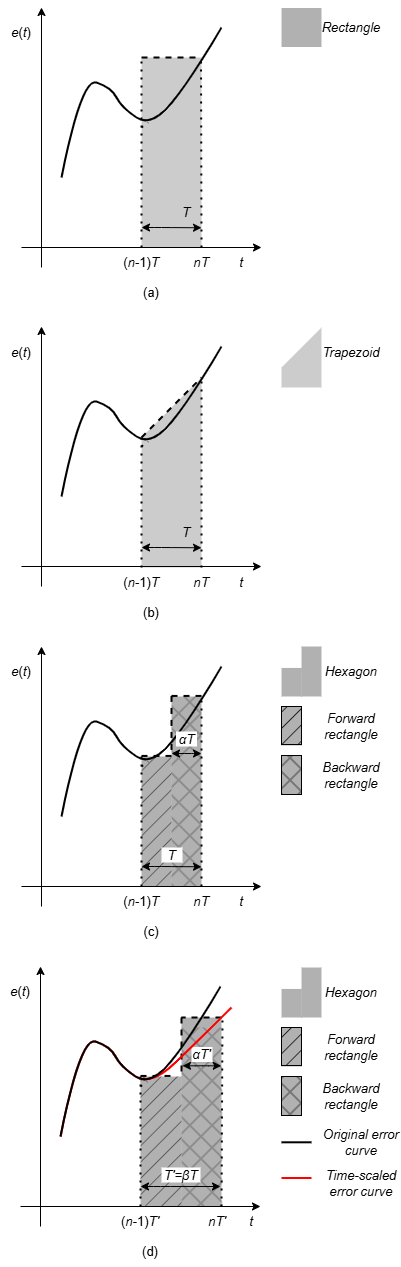}
  \caption{Different approximation approaches based on the numerical integration. 
  (a) Rectangular approximation. (b) Trapezoidal approximation. (c) Hexagonal approximation. (d) Scalable Hexagonal Approximation (SHA).}
  \label{fig_approximation}
\end{figure}

In the continuous domain, the original function ($u(t)$) and the error function ($e(t)$) are expressed as follows: 
\begin{equation} u(t) = \int e(t) \,dt  \label{eq_u_e_cont}\end{equation}
\begin{equation} e(t) = \frac{d u(t)}{d t} \label{eq_diff}\end{equation}

In the discrete domain, $u(n)$ is expressed as follows:
\begin{equation} u(n) = \int_{(n-1)T}^{nT} e(t) \,dt + u(n-1) \label{eq_u_e_disc}\end{equation}

Fig.~\ref{fig_approximation} compares the mathematical interpretation of the SBT and other methods.
Fig.~\ref{fig_approximation}(a) illustrates the rectangular approximation geometrically.
The area enclosed by the three dotted lines and the horizontal axis forms a rectangle.
The rectangular approximation is equivalent to the numerical integration of the rectangular area.
Fig.~\ref{fig_approximation}(b) illustrates the trapezoidal approximation geometrically.
The area enclosed by the three dotted lines and the horizontal axis forms a trapezoid.
The trapezoidal approximation is equivalent to the numerical integration of the trapezoidal area.

Fig.~\ref{fig_approximation}(c) illustrates the hexagonal approximation geometrically.
The area enclosed by the five dotted lines and the horizontal axis forms a hexagon.
This hexagonal approximation is equivalent to the numerical integration of the hexagonal area.
In this case, $e(t)$ is expressed as follows:
\begin{equation} 
  e(t) = 
  \begin{cases}
      e(n-1), & t \in [(n-1)T, (n-\alpha)T] \\
      e(n), & t \in ((n-\alpha)T, n\cdot T]
  \end{cases}
  \label{eq_hexa_error}
\end{equation}
In the discrete domain, $u(n)$ is expressed as follows:
\begin{equation} u(n) = (1-\alpha)\cdot e(n-1)T + \alpha \cdot e(n)T + u(n-1) \label{eq_u_hexa_appr}\end{equation}
Therefore,
\begin{equation} (1 - z^{-1})\cdot U(z) = [(1-\alpha)\cdot z^{-1}\cdot T + \alpha\cdot T] \cdot E(z)\label{eq_Uz_Ez}\end{equation}
\begin{equation} s = \frac{E(z)}{U(z)} = \frac{1}{T}\cdot \frac{z-1}{\alpha z + (1-\alpha)}\label{eq_gbt}\end{equation}

Moreover, the hexagonal area in Fig.~\ref{fig_approximation}(c) comprises two rectangles.
The left part is a forward rectangle and the right part is a backward rectangle.
The physical meaning of the parameter $\alpha$ is the percentage of the backward rectangular area.
Here, we define this factor as shape factor, and denote it as follows:
\begin{equation} \alpha = \frac{S_{bw\_rec}}{S_{bw\_rec} + S_{fw\_rec}}\label{eq_alpha_def}\end{equation}
where $S_{bw\_rec}$ and $S_{fw\_rec}$ are the backward rectangular area and the forward rectangular area, respectively.

Fig.~\ref{fig_approximation}(d) illustrates the scalable hexagonal approximation (SHA) geometrically.
Compared with the hexagonal approximation, SHA introduces a new factor $\beta$ by scaling the sampling time from $T$ to $T'$.
Thus, the original error curve (black solid line) transforms into the time-scaled error curve (red solid line).
Here, we define $\beta$ as time factor, and the physical meaning of $\beta$ is defined as:
\begin{equation} \beta = \frac{T'}{T}\label{eq_beta_def}\end{equation}
where $T$ and $T'$ are the original and scaled sampling time, respectively.
Therefore, the SBT is denoted as:
\begin{equation} s = \frac{1}{\beta \cdot T}\cdot \frac{z-1}{\alpha z+ (1-\alpha)}\label{eq_sbt}\end{equation}

\subsection{Mapping and Stability}
\begin{figure}[htbp] 
  \centering
  \includegraphics[width=1.0\linewidth]{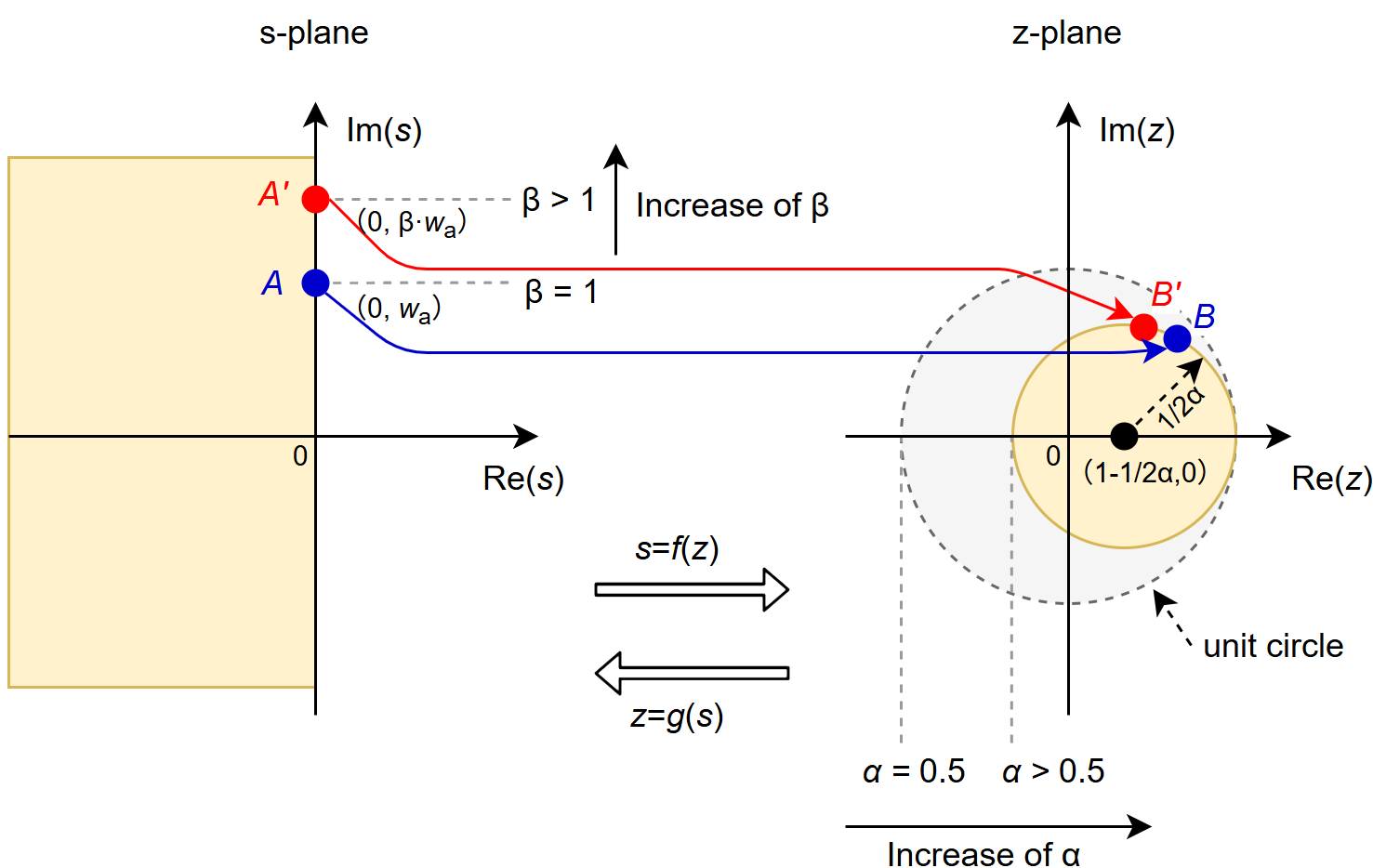}
  \caption{Mapping of s-plane to z-plane}
  \label{fig_s2z_mapping}
\end{figure}

Let $s=\sigma_s + j\omega_s$ and $z=\gamma_z + j\cdot \zeta_z $, we have the following expression by substituting these into equation \eqref{eq_gbt}:
\begin{equation}   
  \begin{split}
      s &= \sigma_s + j\omega_s = \frac{1}{\beta T}\frac{[\gamma_z + j\cdot \zeta_z]-1}{\alpha (\gamma_z + j\cdot \zeta_z)+(1-\alpha)} \\  
         &=\frac{1}{\beta T}\frac{[\alpha(\gamma_z-1)^2+\gamma_z-1+\alpha\cdot \zeta_z^2]+j\alpha\cdot \zeta_z}{[\alpha \gamma_z+1-\alpha]^2+[\alpha\cdot \zeta_z]^2}    \\    
  \end{split}
  \label{eq_stab1}
\end{equation}
Therefore, $\sigma_s$ and $\omega_s$ can be expressed as follows:
\begin{equation} \sigma_s = \frac{1}{\beta T}\frac{\alpha(\gamma_z-1)^2+\gamma_z-1+\alpha\cdot \zeta_z^2}{[\alpha \gamma_z+1-\alpha]^2+[\alpha\cdot \zeta_z]^2} \label{eq_sigma_s}\end{equation}
\begin{equation} \omega_s = \frac{1}{\beta T}\frac{\zeta_z}{[\alpha \gamma_z+1-\alpha]^2+[\alpha\cdot \zeta_z]^2} \label{eq_omega_s}\end{equation}
The transformation is stable unless the left half of the s-plane is mapped within the unit circle of the z-plane, 
which implies that the stable constraint is $\sigma_s \leq  0$.
Substituting this into equation \eqref{eq_sigma_s} yields,
\begin{equation} [\gamma_z - (1-\frac{1}{2\alpha})]^2 + \zeta_z^2 \leq (\frac{1}{2\alpha})^2\label{eq_stab_cond}\end{equation}
In the z-plane, the image of the mapping is a circle. 
As shown in Fig.~\ref{fig_s2z_mapping}, its center is located at $(1-\frac{1}{2\alpha}, 0)$ and its radius is $\frac{1}{2\alpha}$.
As $\alpha$ increases, the mapped image becomes smaller.
Fig.~\ref{fig_s2z_mapping} also illustrates how the mapping changes when adjusting $\beta$.
Increasing $\beta$ shifts the mapped point in the z-plane from point $B$ (which corresponds to point $A$ in the s-plane) to 
point $B'$ (which corresponds to point $A'$ in the s-plane).
The restriction for a stable transformation in z-plane can be expressed as follows:
\begin{equation} 1-\frac{1}{2\alpha}-\frac{1}{2\alpha} \geq -1 \label{eq_stab_cond1}\end{equation}
yields $\alpha \geq 0.5$.
In conclusion, the shape factor $\alpha$ should be within the range [0.5, 1] in order to achieve a stable transformation.

\subsection{Relations with the Existing Methods}
\begin{table*}[!t] 
\caption{Relations with the Existing Methods\label{tab:relation}}
\centering
\begin{tabular}{|c|c|c|c|c|}
\hline
\textbf{Method} & \textbf{Transformation} & \textbf{Parameters} & \textbf{Relationship} & \textbf{Notes} \\
\hline
SBT                            & $s = \frac{1}{\beta\cdot T}\frac{z-1}{\alpha z+(1-\alpha)}$        & $\alpha \in [0,1]$, $\beta \in (0, \infty)$ & \textbf{reference} & universal form of the bilinear transformation\\
GBT in \cite{GBT-01-Sekara} -- \cite{GBT-02}   & $s = \frac{1}{T}\frac{z-1}{\alpha z+(1-\alpha)}$        & $\alpha \in [0,1]$                          & $\beta=1$ & a specific case of the SBT\\
Euler method                   & $s = \frac{1}{T}\frac{z-1}{z}$                          & /                                           & $\alpha=1, \beta=1$   & a specific case of the SBT\\
Tustin method                  & $s = \frac{2}{T}\frac{z-1}{z+1}$                        & /                                           & $\alpha=0.5, \beta=1$ & a specific case of the SBT\\
Tustin with pre-warping        & $s = \frac{2}{T}\frac{z-1}{z+1}$                        & $K_{pw}=\frac{w_{pw}}{w_n}$                 & $\alpha=0.5, \beta=K_{pw}$ & similar performance with the SBT under this param. set\\
\hline
\end{tabular}
\end{table*}

Table~\ref{tab:relation} shows the relationship between the SBT and some existing methods. 
Compared with the GBT method \cite{GBT-01-Sekara} -- \cite{GBT-02}, the SBT introduces an additional degree of freedom, 
the time factor $\beta$. 
The Euler method constitutes a specific case of the SBT framework, obtained when $\alpha = 1$ and $\beta = 1$.
Similarly, the Tustin method is another specific instance of the SBT, 
corresponding to the parameter set $\alpha = 0.5$ and $\beta = 1$.
The pre-warped Tustin method can be viewed as an enhancement of the standard Tustin method, 
achieved by changing the original frequency from $w_{n}$ to a pre-warped value $w_{pw}$. 
The pre-warped factor is defined as:
\begin{equation}
  K_{pw} = \frac{w_{pw}}{w_{n}}\label{eq_k_pw}
\end{equation}
where $w_{n}$ and $w_{pw}$ denote the original and pre-warped frequencies, respectively.
It should be noted that the SBT method can yield similar performance as the Tustin with pre-warping 
when setting $\alpha$ to 0.5, and $\beta$ from 1 to a pre-warped factor $K_{pw}$.

\section{Application in Quasi-Resonant Controller}\label{sec3}
\subsection{QR Controllers in the Continuous Domain}\label{sub1sec3}
The continuous-time QR controller can be expressed as a second-order transfer function in the $s$-domain:
\begin{equation}
  G_{QR}(s) = \frac{2K_r\cdot \omega_c\cdot s}{s^2 + 2\omega_c\cdot s + \omega_n^2}\label{eq_cont_qr}
\end{equation}
where $\omega_n$ denotes the resonant angular frequency, 
$\omega_c$ represents the cutoff bandwidth which broadens the resonant peak and enhances robustness against minor frequency variations, 
and $K_r$ represents the resonant gain determining the peak gain at $\omega_n$.

\subsection{QR Controllers in the Discrete Domain}\label{sub2sec3}
The general transfer function of the QR controller in the $z$-domain can be expressed as:
\begin{equation}
  G_{QR}(z) = \frac{a_2\cdot z^2 + a_1\cdot z + a_0}{b_2\cdot z^2 + b_1\cdot z + b_0}\label{eq_disc_qr}
\end{equation}
where $a_i (i \in \{0,1,2\})$ and $b_j (j \in \{0,1,2\})$ denote the numerator and denominator coefficients, respectively.

As shown in Table~\ref{tab:discQR}, 
the coefficients of the four discrete QR controllers derived from different discretization methods are presented. 
For a given continuous QR controller, the parameters $K_r$, $\omega_c$, and $\omega_n$ are treated as constants during the discretization process.
It is evident that the Euler and Tustin methods do not introduce any additional variables. 
However, for the Tustin method with pre-warping, the pre-warped factor $K_{pw}$ must be calculated using the following expression \cite{GBT-03}:
\begin{equation}
  K_{pw} = \frac{\tan(\frac{\omega_n T}{2})}{\frac{\omega_n T}{2}} \label{eq_Kpw}
\end{equation}
Unlike these conventional approaches, 
the proposed SBT method is characterized by two distinct degrees of freedom. 
In practice, there are generally two approaches to designing these parameters: the straightforward approach and the optimal approach.
The straightforward approach is depicted as follows:
\begin{equation}
  (\alpha,\beta) =(0.5,K_{pw})\label{eq_simple_approach}
\end{equation}
The optimal approach is depicted as follows:
\begin{equation} 
    \begin{split}
    \min_{(\alpha,\beta)}   \quad &Q_{loss}(\alpha,\beta)\\
    \text{subject to}       \quad &0.5 \leq \alpha \leq 1, \beta > 0 \\
    \end{split}
    \label{eq_optimal_approach}
\end{equation}
where $Q_{loss}(\alpha,\beta)$ is the loss function with $\alpha$ and $\beta$ as independent variables.

In this study, we adopt the straightforward approach for its ease of implementation and its similarity to the SOTA method. 
However, a key distinction lies in the parameter regulation mechanism: 
the proposed SBT method simultaneously regulates the $\omega_n$ and $\omega_c$, 
whereas the SOTA method solely regulates the $\omega_n$. 
This modification yields discernible performance enhancements, as confirmed by the experimental results.

\begin{table*}[!t] 
\caption{Coefficients of Discrete Transfer Function Implemented by Different Discretization Methods\label{tab:discQR}}
\centering
\begin{tabular}{|c|c|c|c|c|}
\hline
\textbf{Param.} & \textbf{Euler} & \textbf{Tustin} & \textbf{Tustin with pre-warping} & \textbf{Our Method (SBT)}\\
\hline
$a_2$ & $2K_r\cdot\omega_c\cdot T$                 & $K_r\cdot \omega_c\cdot T$                 & $K_r\cdot \omega_c\cdot T$                             & $2\alpha\cdot \beta\cdot K_r\cdot \omega_c\cdot T$ \\
$a_1$ & $-2K_r\cdot\omega_c\cdot T$                & 0                                          & 0                                                      & $-(4\alpha-2)\beta\cdot K_r\cdot \omega_c\cdot T$ \\
$a_0$ & 0                                          & $-K_r\cdot \omega_c\cdot T$                & $-K_r\cdot \omega_c\cdot T$                            & $-(2-2\alpha)\beta\cdot K_r\cdot \omega_c\cdot T$ \\
$b_2$ & $1+2\omega_c\cdot T + (\omega_n\cdot T)^2$ & $1+\omega_c\cdot T+(0.5\omega_n\cdot T)^2$ & $1+\omega_c\cdot T+(0.5K_{pw}\cdot \omega_n\cdot T)^2$ & $1+2\alpha\cdot\beta\cdot\omega_c\cdot T+(\alpha\cdot\beta\cdot\omega_n\cdot T)^2$ \\
$b_1$ & $-2-2\omega_c\cdot T$                      & $0.5(\omega_n\cdot T)^2-2$                 & $0.5(K_{pw}\cdot \omega_n\cdot T)^2-2$                 & $-2-(4\alpha-2)\beta\cdot\omega_c\cdot T+2\alpha(1-\alpha)(\beta\cdot\omega_n\cdot T)^2$ \\
$b_0$ & 1                                          & $1-\omega_c\cdot T+(0.5\omega_n\cdot T)^2$ & $1-\omega_c\cdot T+(0.5K_{pw}\cdot \omega_n\cdot T)^2$ & $1-(2-2\alpha)\beta\cdot\omega_c\cdot T+[(1-\alpha)\beta\cdot\omega_n\cdot T]^2$\\
\hline
\end{tabular}
\end{table*}

\subsection{Frequency Response of the Discrete QR Controllers}\label{sub3sec3}
\begin{figure}[htbp] 
  \centering
  \includegraphics[width=1.0\linewidth]{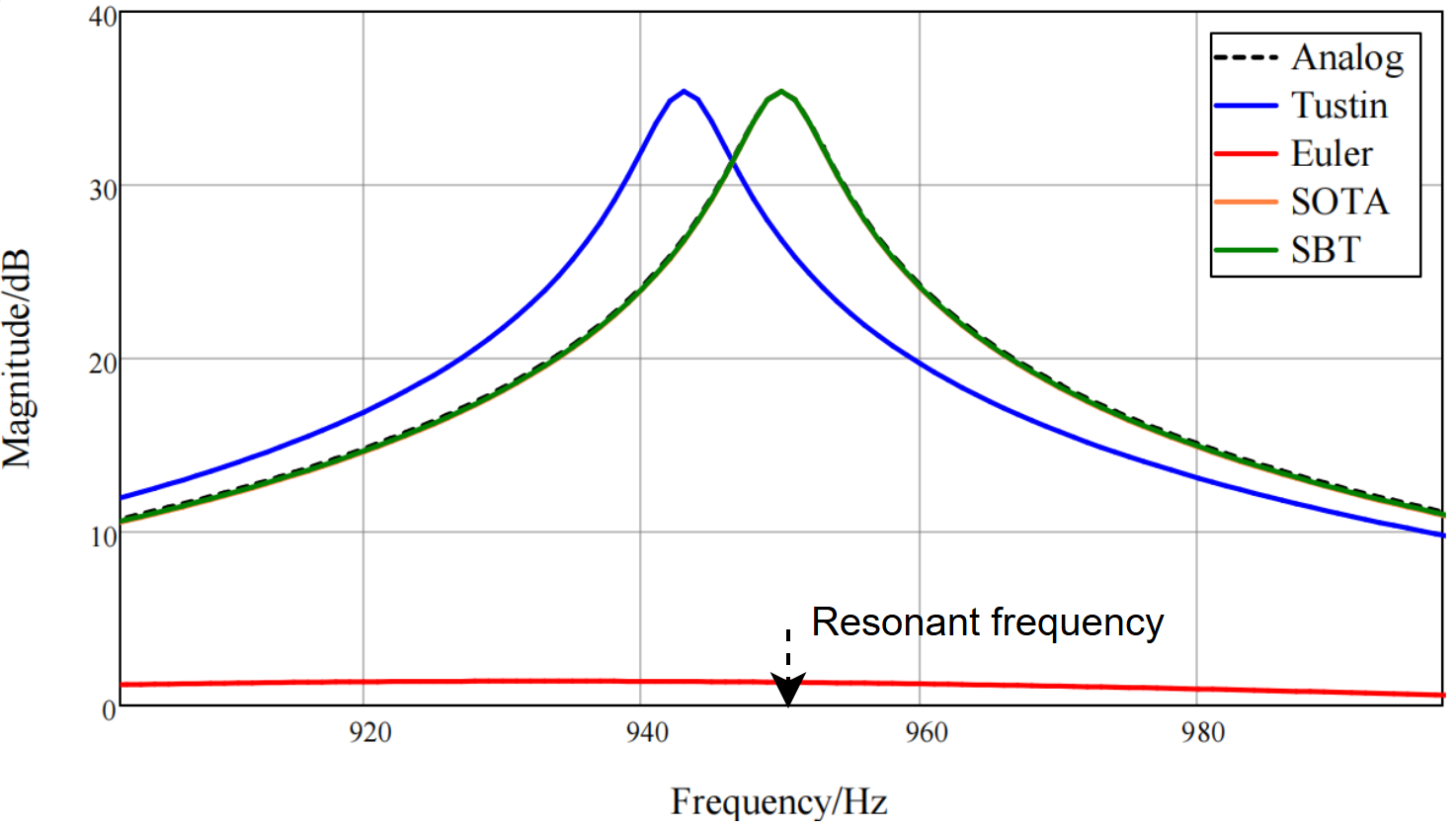}
  \caption{Frequency response of the QR controllers}
  \label{fig_freq_resp}
\end{figure}

Fig.~\ref{fig_freq_resp} compares the magnitude responses of the analog QR controller 
and its four discrete counterparts obtained via the Euler, Tustin, SOTA, and proposed SBT methods. 
The analog controller is depicted as a black dashed line, 
whereas the discrete versions (Euler, Tustin, SOTA, and SBT) are represented by red, blue, orange, and green solid lines, respectively.
Significant resonance damping is observed in the Euler discretization, 
whereas the Tustin method suffers from noticeable frequency warping. 
In contrast, both the SOTA and the proposed SBT methods demonstrate superior performance, 
with magnitude characteristic closely aligning with that of the analog controller.

\begin{figure}[htbp] 
  \centering
  \includegraphics[width=1.0\linewidth]{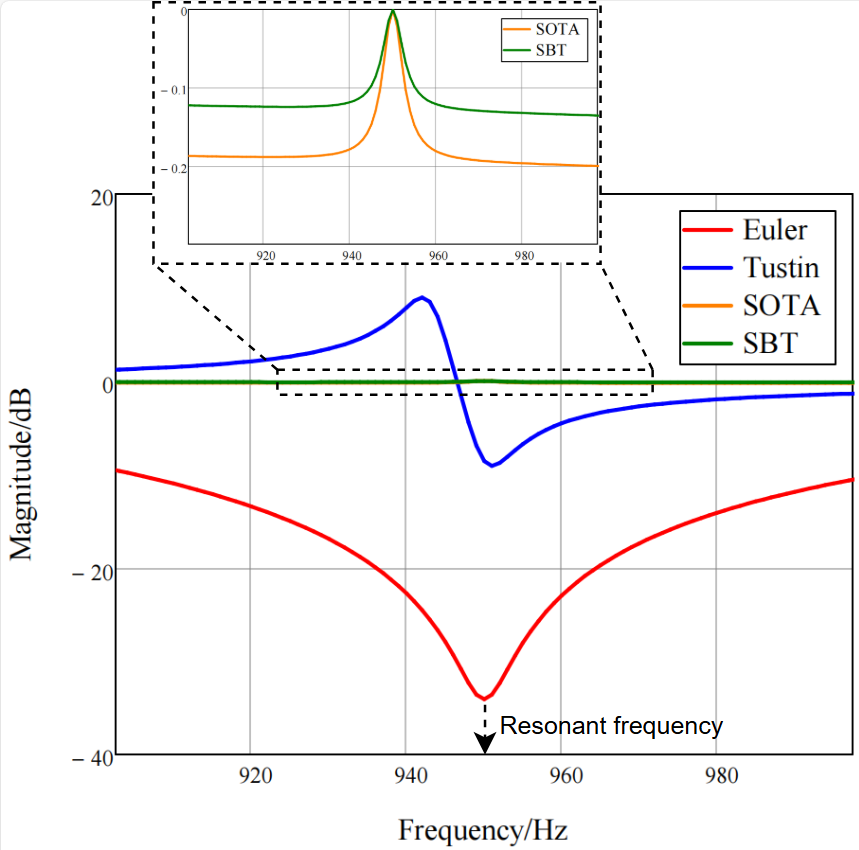}
  \caption{Magnitude error under different discretization methods}
  \label{fig_mag_err}
\end{figure}

Fig.~\ref{fig_mag_err} illustrates the magnitude error curves of the discrete QR controllers 
derived via the Euler, Tustin, SOTA, and proposed SBT methods, indicated by red, blue, orange, and green solid lines, respectively. 
The magnitude error is defined as the difference obtained by subtracting the discrete magnitude from the analog counterpart. 
It is evident that the Euler and Tustin methods introduce significant errors near the resonant frequency. 
In contrast, the errors associated with both the SOTA and the proposed SBT methods are negligible. 
Furthermore, a magnified view of the error curves confirms that the proposed SBT method yields a smaller error than the SOTA method.

\subsection{Theoretical Explanation on the Discretization Error}\label{sub4sec3}
The location of the poles determines the frequency response of the QR controllers.
In this study, we use pole mapping in order to explain the discretization error as shown in Fig.~\ref{fig_mag_err}.
The continuous QR controller has two original poles in the $s$-plane, which are $(\sigma_0, \omega_0)$, and $(\sigma_0, -\omega_0)$, respectively.
Here, $\sigma_0$ and $\omega_0$ denote the real and imaginary parts of the original pole, respectively, and are depicted as follows:
\begin{equation}
  \begin{cases}
    \sigma_0 = -\omega_c\\
    \omega_0 = \sqrt{\omega_n^2-\omega_c^2}
  \end{cases}
  \label{eq_ori_pole}
\end{equation}
To illustrate the derivation process, we take the pole $(\sigma_0, \omega_0)$ under the Tustin method as an example. 
The corresponding mapped pole in the $z$-plane is derived as follows:
\begin{equation}
    z=\frac{2+s T}{2-s T} = \frac{2+(\sigma_0+j\omega_0)T}{2-(\sigma_0+j\omega_0)T} = \frac{4-(\omega_n T)^2+j4\omega_0 T}{(2-\sigma_0 T)^2+(\omega_0 T)^2}
  \label{eq_mapped_pole_tustin}
\end{equation}
Consequently, the coordinates $(\gamma_{tustin}, \zeta_{tustin})$ of the mapped pole in the $z$-plane are defined as:
\begin{equation}
  \begin{cases}
    \gamma_{tustin} = \frac{4-(\omega_n T)^2}{(2-\sigma_0 T)^2+(\omega_0 T)^2}\\
    \zeta_{tustin} = \frac{4\omega_0 T}{(2-\sigma_0 T)^2+(\omega_0 T)^2}
  \end{cases}
  \label{eq_mapped_pole_tustin1}
\end{equation}
where $\gamma_{tustin}$ and $\zeta_{tustin}$ represent the real and imaginary parts of the mapped pole in the $z$-plane, respectively.

This pole, located in the $z$-plane, is transformed back to the $s$-plane using the inverse of the exact discretization mapping ($s = \ln(z)/T$).
The equivalent pole via the Tustin discretization is then derived as follows:
\begin{equation}
  \begin{cases}
    \sigma_{tustin} = \frac{1}{T}\ln\left(\frac{4\omega_0 T}{[(2-\sigma_0 T)^2+(\omega_0 T)^2]\sin\left(\arctan\left(\frac{4\omega_0 T}{4-(\omega_n T)^2}\right)\right)}\right)\\
    \omega_{tustin} = \frac{1}{T}\arctan\left(\frac{4\omega_0 T}{4-(\omega_n T)^2}\right)
  \end{cases}
  \label{eq_equivalent_pole_tustin}
\end{equation}
where $\sigma_{tustin}$ and $\omega_{tustin}$ represent the real and imaginary parts of the equivalent pole in the $s$-plane, respectively.

For other discretization methods, including the Euler, SOTA, and SBT ($\alpha=0.5, \beta=K_{pw}$) methods, 
the mapped poles in the $z$-plane and equivalent poles in the $s$-plane are summarized in Table~\ref{tab:pole_mapping}.
In this table, $\sigma_1$ and $\omega_1$ denote the real and imaginary parts of the pre-warped pole in the $s$-plane, defined as follows:
\begin{equation}
  \begin{cases}
    \sigma_1 = -\omega_c\\
    \omega_1 = \sqrt{(K_{pw}\omega_n)^2-\omega_c^2}
  \end{cases}
  \label{eq_pw_pole}
\end{equation}

\begin{table*}[!t] 
\caption{Mapping of Pole with Different Discretization Methods\label{tab:pole_mapping}}
\centering
\begin{tabular}{|c|c|c|}
\hline
\textbf{Method} & \textbf{Mapped Pole in the $z$-plane} & \textbf{Equivalent Pole in the $s$-plane} \\
\hline
Euler  & $\left(\frac{1-\sigma_0 T}{(1-\sigma_0 T)^2+(\omega_0 T)^2}, \frac{\omega_0 T}{(1-\sigma_0 T)^2+(\omega_0 T)^2}\right)$                                      & $\left(\frac{1}{T}ln\left(\frac{\omega_0 T}{[(1-\sigma_0 T)^2+(\omega_0 T)^2]sin\left(arctan(\frac{\omega_0 T}{1-\sigma_0 T})\right)}\right), \frac{1}{T}arctan\left(\frac{\omega_0 T}{1-\sigma_0 T}\right)\right)$ \\
Tustin & $\left(\frac{4-(\omega_n T)^2}{(2-\sigma_0 T)^2+(\omega_0 T)^2}, \frac{4\omega_0 T}{(2-\sigma_0 T)^2+(\omega_0 T)^2}\right)$                                 & $\left(\frac{1}{T}ln\left(\frac{4\omega_0 T}{[(2-\sigma_0 T)^2+(\omega_0 T)^2]sin\left(arctan(\frac{4\omega_0 T}{4-(\omega_n T)^2})\right)}\right), \frac{1}{T}arctan\left(\frac{4\omega_0 T}{4-(\omega_n T)^2}\right)\right)$ \\
SOTA   & $\left(\frac{4-(K_{pw}\omega_n T)^2}{(2-\sigma_1 T)^2+(\omega_1 T)^2}, \frac{4\omega_1 T}{(2-\sigma_1 T)^2+(\omega_1 T)^2}\right)$                           & $\left(\frac{1}{T}ln\left(\frac{4\omega_1 T}{[(2-\sigma_1 T)^2+(\omega_1 T)^2]sin\left(arctan(\frac{4\omega_1 T}{4-(K_{pw}\omega_n T)^2})\right)}\right), \frac{1}{T}arctan\left(\frac{4\omega_1 T}{4-(K_{pw}\omega_n T)^2}\right)\right)$ \\
SBT    & $\left(\frac{4-(\beta\omega_n T)^2}{(2-\beta\sigma_0 T)^2+(\beta\omega_0 T)^2}, \frac{4\beta\omega_0 T}{(2-\beta\sigma_0 T)^2+(\beta\omega_0 T)^2}\right)$   & $\left(\frac{1}{T}ln\left(\frac{4\beta\omega_0 T}{[(2-\beta\sigma_0 T)^2+(\beta\omega_0 T)^2]sin\left(arctan(\frac{4\beta\omega_0 T}{4-(\beta\omega_n T)^2})\right)}\right), \frac{1}{T}arctan\left(\frac{4\beta\omega_0 T}{4-(\beta\omega_n T)^2}\right)\right)$ \\
\hline
\end{tabular}
\end{table*}

This study investigates a QR controller parameterized with $\omega_n=5969, \omega_c=17.907, K_r=59.1$. 
The pole mappings in the $z$-plane and their equivalent s-plane representations under different discretization methods are summarized in Table~\ref{tab:pole_mapping_case},
where the exact discretization method serves as the reference benchmark.
The Euler method induces considerable distortion in the real part ($\sigma$) of the pole, 
which accounts for the pronounced resonance damping observed in the frequency response.
In contrast, the Tustin method introduces significant inaccuracy in the imaginary part ($\omega$), 
leading to a noticeable frequency warping effect. 
Among the compared methods, the SOTA and SBT method both perform well, 
with the mapped and equivalent poles closely aligned with the reference. 
Nevertheless, the proposed SBT method demonstrates superior accuracy,
exhibiting the smallest deviation between mapped and equivalent poles relative to the reference.

\begin{figure*}[htbp] 
  \centering
  \includegraphics[width=0.87\linewidth]{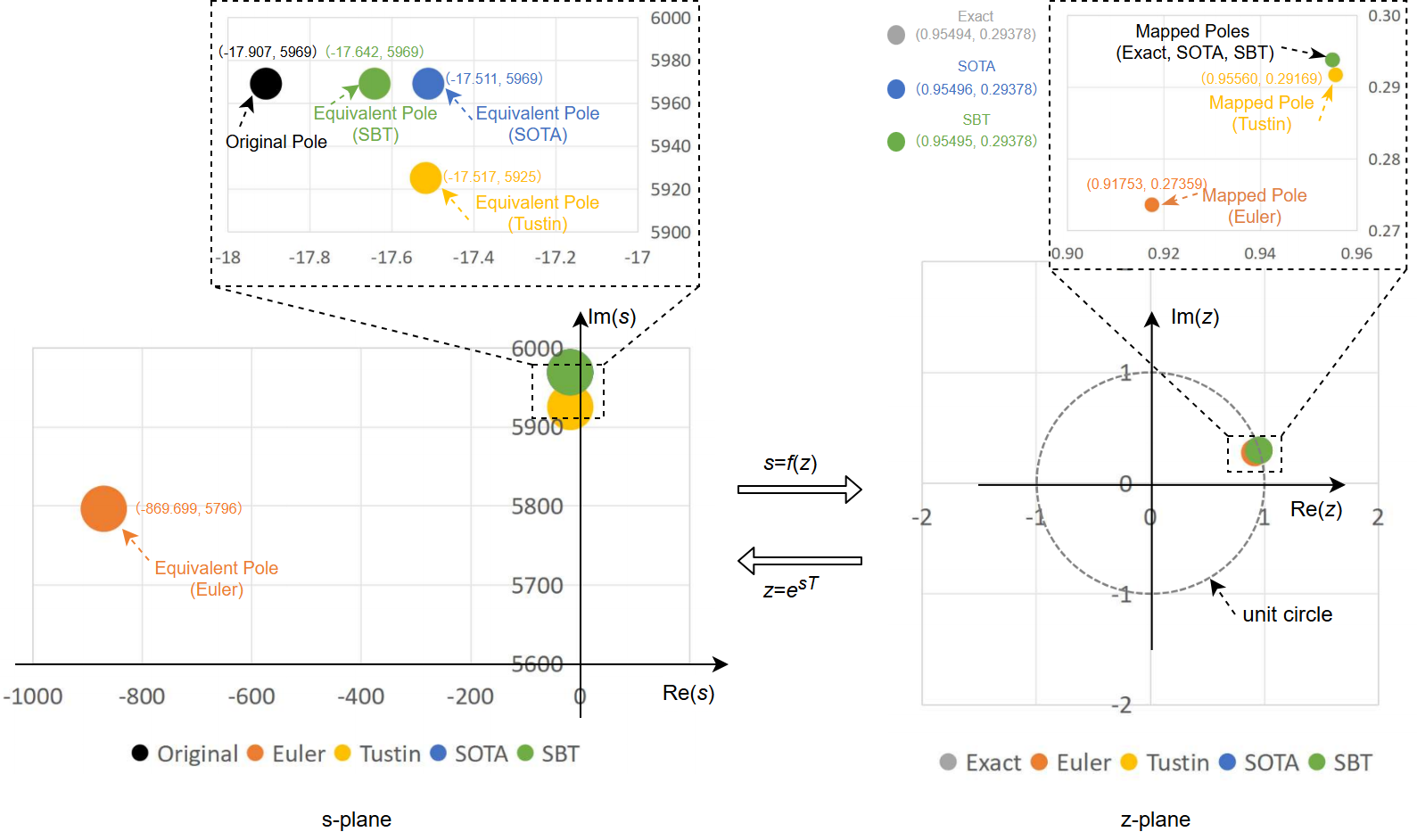}
  \caption{Mapping Graph of Poles under different discretization methods}
  \label{fig_proof}
\end{figure*}

The pole mapping under various discretization methods is illustrated in Fig.~\ref{fig_proof}. 
The original continuous-time pole is denoted by a black solid circle, 
while the discrete counterparts—Exact, Euler, Tustin, SOTA, and SBT—are represented by gray, orange, yellow, blue, and green solid circles, respectively.
It is evident that the equivalent pole obtained via the SBT method demonstrates superior performance, 
exhibiting a near-perfect alignment with the original pole in the $s$-plane. 
Consequently, the theoretical explanation is effectively verified.

\begin{table}[htbp] 
\caption{A Case of Pole Mapping with Different Discretization Methods\label{tab:pole_mapping_case}}
\centering
\begin{tabular}{|c|c|c|c|}
\hline
\textbf{Method} & \textbf{Mapped Pole ($z$)} & \textbf{Equivalent Pole ($s$)} & \textbf{Notes}\\
\hline
Exact  & (0.95494, 0.29378) & (-17.907, 5969) & \textbf{Ref.}\\
Euler  & (0.91753, 0.27359) & (-869.699, 5796)& $\sigma$ warps\\
Tustin & (0.95560, 0.29169) & (-17.517, 5925) & $\omega$ warps\\
SOTA   & (0.95496, 0.29378) & (-17.511, 5969) & close to Ref.\\
SBT    & (0.95495, 0.29378) & (-17.642, 5969) & closet to Ref.\\
\hline
\end{tabular}
\end{table}

\section{Experimental Evaluation}\label{sec4}
\subsection{Experiments on a Control Board}\label{sub1sec4}
\begin{figure}[htbp]  
  \centering
  \includegraphics[width=1.0\linewidth]{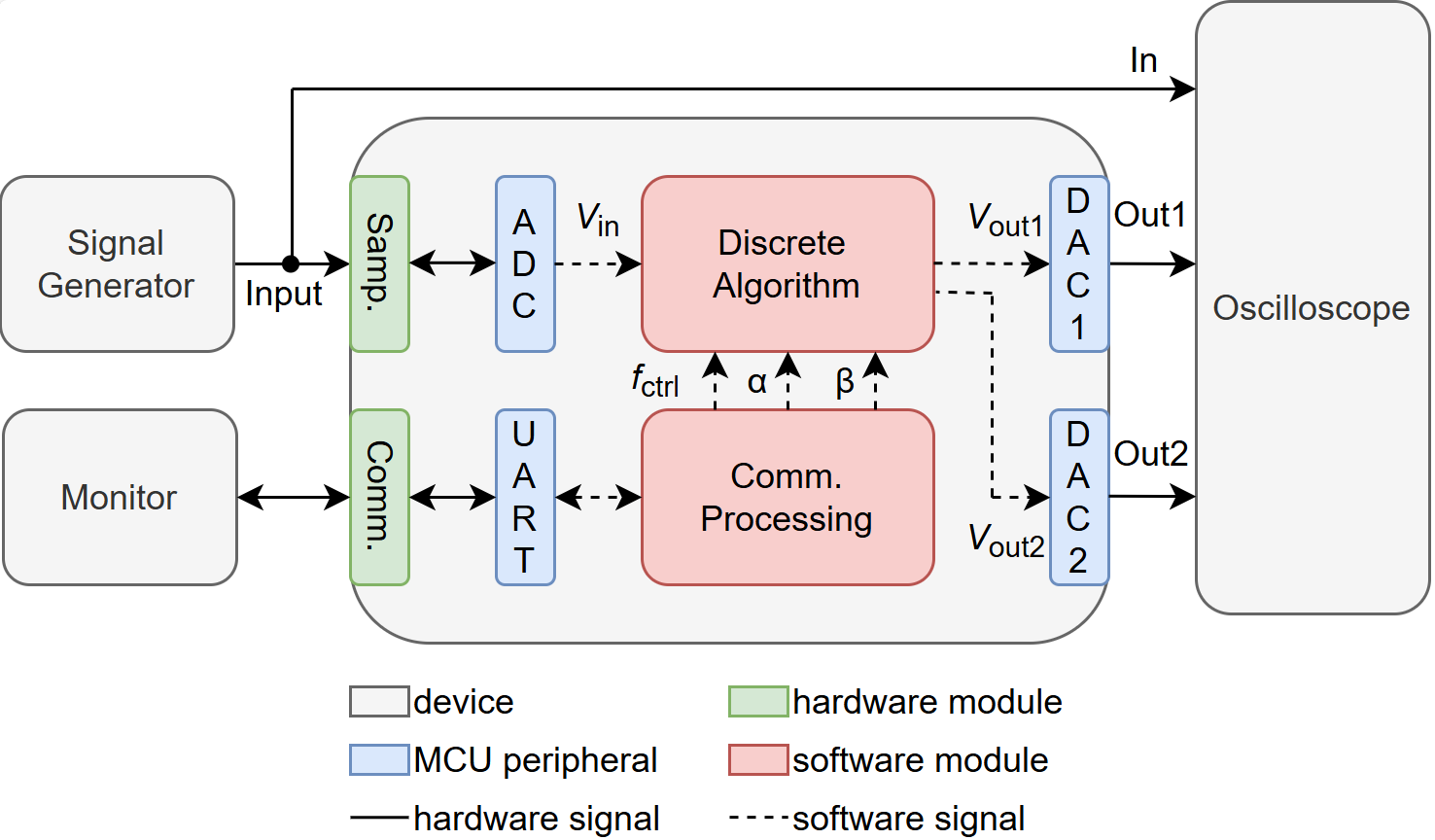}
  \caption{Block diagram of the control board test. Samp.: Sampling, Comm.: Communication, ADC: A/D Conversion, DAC: D/A Conversion.}  
  \label{fig_board_test_diagram}
\end{figure}

\begin{figure}[htbp]  
  \centering
  \includegraphics[width=0.8\linewidth]{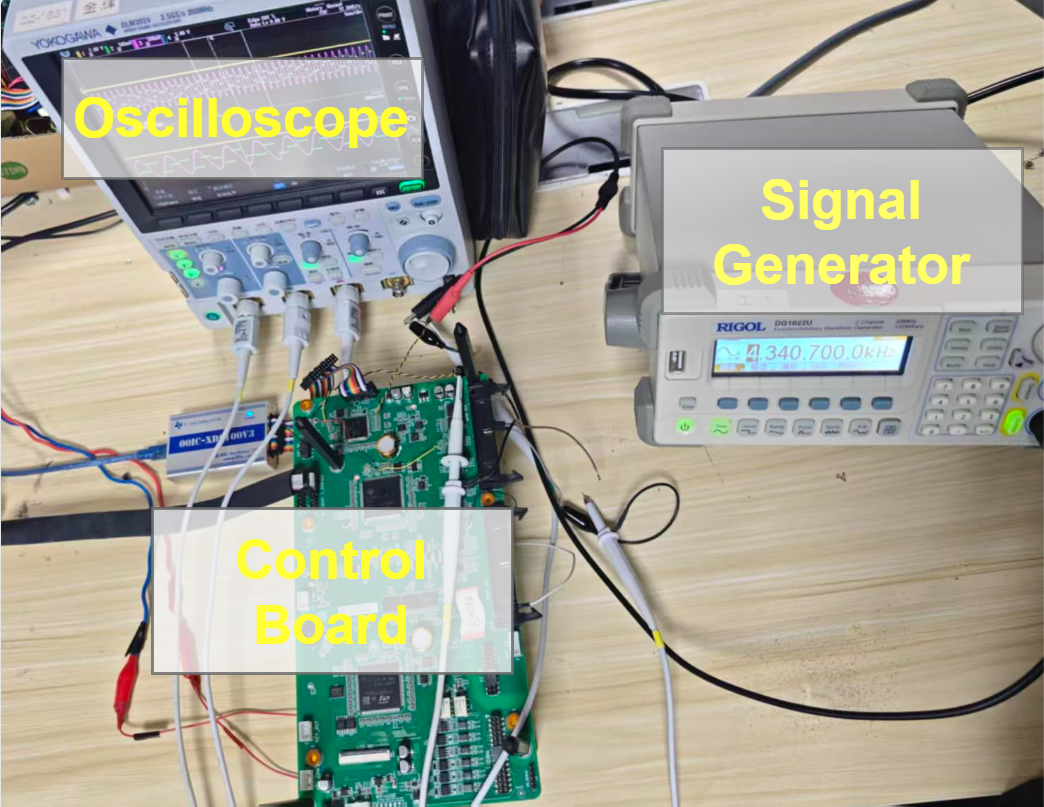}
  \caption{Physical setup of the control board test.}
  \label{fig_board_test_setup}
\end{figure}

To validate the effectiveness of the proposed SBT,
a discrete QR controller is built as shown in Fig.~\ref{fig_board_test_diagram}.
The algorithms are deployed on a control board equipped with a TMS320F28P65 microcontroller.
A high-precision signal generator, DG1022U, provides the input signal.
A laptop serves as a monitoring terminal and communicates with the control board via RS485 to configure parameters,
including the sampling frequency ($f_{samp}$), the shape factor ($\alpha$), and the time factor ($\beta$).
The "Discrete Algorithm" is the key module which implements the discrete algorithm 
in form of a difference equation illustrated as follows:
\begin{equation}
  \begin{split}
  V_{out}(n) = &K_{in0}\cdot V_{in}(n) + K_{in1}\cdot V_{in}(n-1) + K_{in2}\cdot V_{in}(n-2)\\
               &+ K_{out1}\cdot V_{out}(n-1) + K_{out2}\cdot V_{out}(n-2)\\
  \end{split}  
  \label{eq_qr_diff_form}
\end{equation}
where $V_{X}(n)$, $V_{X}(n-1)$, and $V_{X}(n-2)$(X="in" or "out") are the computed results of the current,
the last sampling period, and the last two sampling period, respectively.
$K_{X0}$, $K_{X1}$, and $K_{X2}$ (X="in" or "out") are the coefficients for 
the $V_{X}(n)$, $V_{X}(n-1)$, and $V_{X}(n-2)$, respectively. 
These coefficients can be calculated according to the equation \eqref{eq_disc_qr}.
When these coefficients are pre-computed outside the main interrupt service routine (ISR), 
the computational time required by each method is identical.
The output signals are sent to the D/A converter (DAC) to be observed in an oscilloscope.
The physical setup of the control board test is shown in Fig.~\ref{fig_board_test_setup}.

\begin{figure*}[htbp]  
  \centering
  \includegraphics[width=1.0\linewidth]{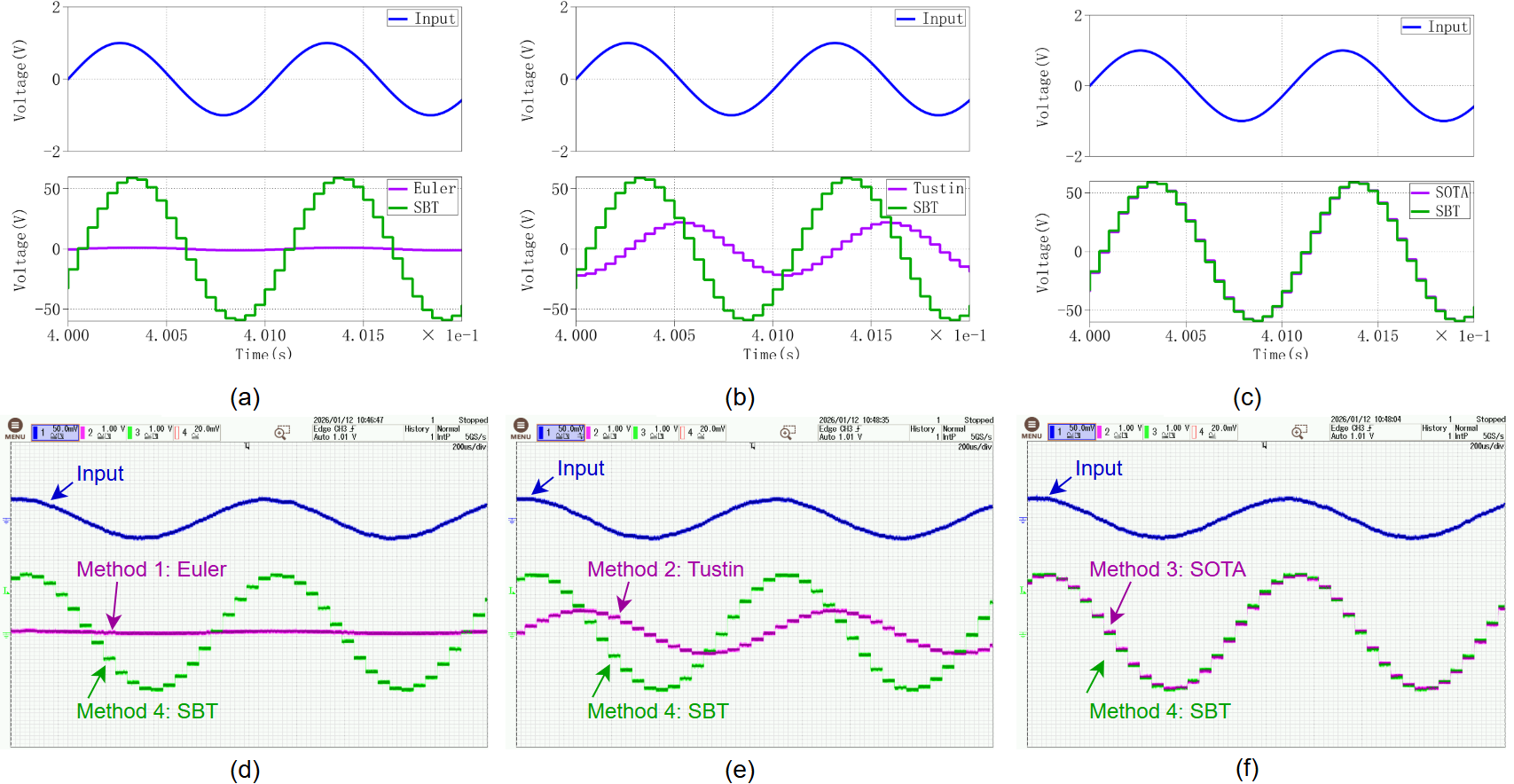}
  \caption{Simulated and experimental waveforms of the control board test. CH1, input voltage, 50 mV/div, CH2, the first output voltage, 1 V/div; CH3, the second output voltage, 1 V/div; time, 200 $\mu$s/div.}
  \label{fig_board_test_waveform}
\end{figure*}

The simulated and experimental waveforms at the resonant frequency are presented in Fig.~\ref{fig_board_test_waveform}. 
Subfigures (a) to (c) display the simulated results obtained via PLECS. 
In these simulations, the amplitude of the input voltage (indicated by the solid pink line) is set to 1 V, 
and the output voltage generated by the proposed SBT method is used as the reference.
The simulation results reveal significant performance variations among the methods. 
The Euler method exhibits substantial resonance damping, 
with an output amplitude of only 1.14 V compared to the reference amplitude of 58.95 V provided by the SBT method. 
The Tustin method demonstrates severe amplitude attenuation and a noticeable phase lag. 
In contrast, the SOTA method performs similarly to the proposed SBT in this case, 
yielding an output amplitude of 58.83 V.
Subfigures (d) to (f) in Fig.~\ref{fig_board_test_waveform} show the corresponding experimental waveforms measured on the control board. 
Here, the input voltage amplitude (solid pink line) is set to 25 mV to ensure the output voltage's peak-to-peak value remains within the 3.3 V limit of the hardware, 
and the SBT's output is again taken as the reference.
The experimental tests corroborate the simulation findings. 
The Euler method again shows pronounced resonance damping, 
with an output amplitude of just 0.0291 V versus the SBT reference of 1.4796 V. 
The Tustin method suffers from significant amplitude damping and phase lag. 
The SOTA method performs closely to the SBT, 
achieving an amplitude of 1.4745 V. 
In conclusion, the experimental results are in strong agreement with the simulations, 
validating the prior theoretical analysis.

\begin{figure}[htbp]  
  \centering
  \includegraphics[width=1.0\linewidth]{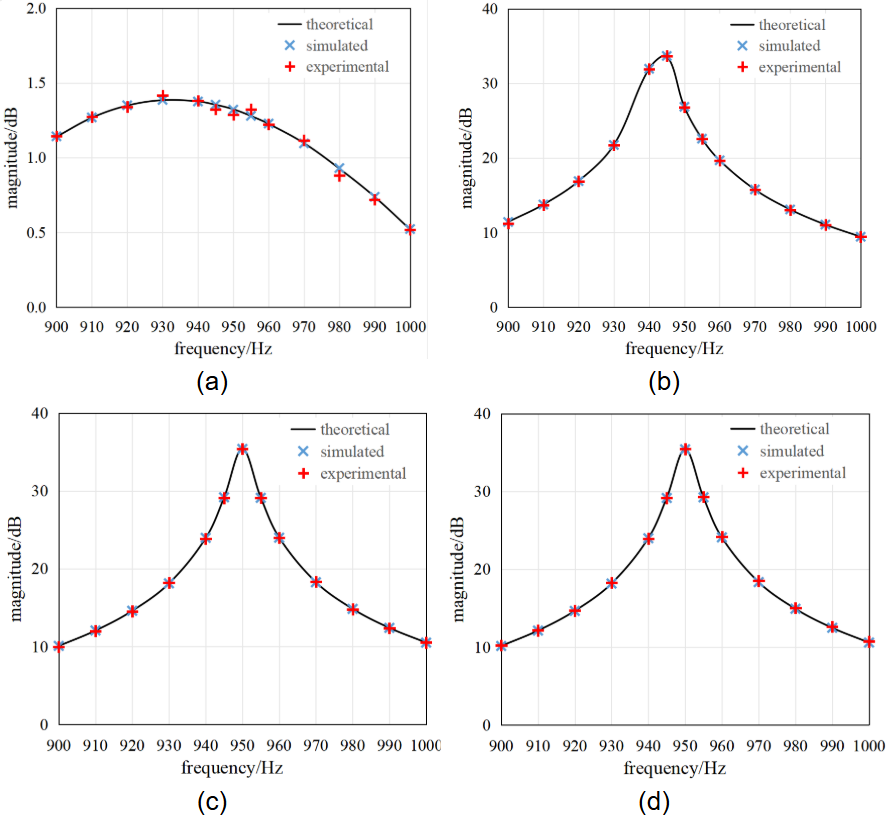}
  \caption{Magnitude curve comparison among theoretical analysis, simulation, and experiments.}
  \label{fig_mag_curve_exp}
\end{figure}

Fig.~\ref{fig_mag_curve_exp} presents a comparative analysis of the frequency responses 
derived from theoretical analysis, simulation, and experimental measurements 
under different discretization methods. 
Subfigures (a) to (d) correspond to the results 
for the Euler, Tustin, SOTA, and the proposed SBT methods, respectively. 
In each subfigure, the theoretical frequency response is plotted as a black solid line, 
while the simulation and experimental results are denoted by blue 'x' and red '+' markers, respectively. 
For all discretization methods, the simulation results show an error rate within 0.5 \%, 
and the experimental error rates remain below 5.0 \% (Tustin, SOTA, and SBT are all within 2.0 \%). 
This close agreement strongly validates the accuracy of the proposed frequency domain model 
described in Table~\ref{tab:discQR}.

\begin{figure}[htbp]  
  \centering
  \includegraphics[width=1.0\linewidth]{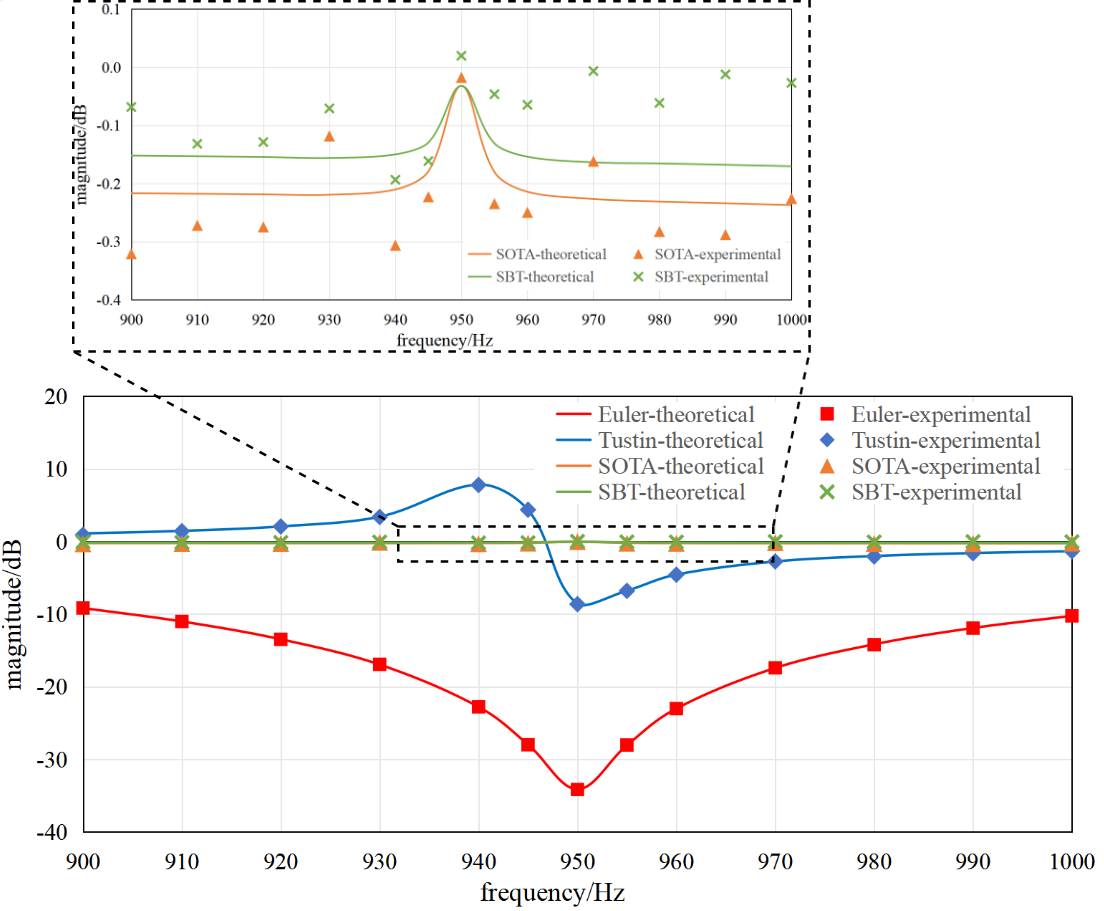}
  \caption{Magnitude error comparison between theoretical analysis and experiments.}
  \label{fig_mag_err_exp}
\end{figure}

A comparison of the experimental magnitude errors for the discrete QR controllers 
derived from the Euler, Tustin, SOTA, and proposed SBT methods is provided in Figure~\ref{fig_mag_err_exp}. 
The solid lines (red, blue, orange, green) correspond to the theoretical error curves for each method, 
while the discrete data points (red, blue, orange, green) represent the experimental measurements. 
Notably, significant errors are observed around the resonant frequency for both the Euler and Tustin methods. 
In contrast, the errors associated with the SOTA and the proposed SBT methods are negligible. 
Moreover, a magnified view of the error profiles reveals that the SBT method yields the smallest error, 
demonstrating higher accuracy than the SOTA method.

\begin{figure}[htbp]  
  \centering
  \includegraphics[width=1.0\linewidth]{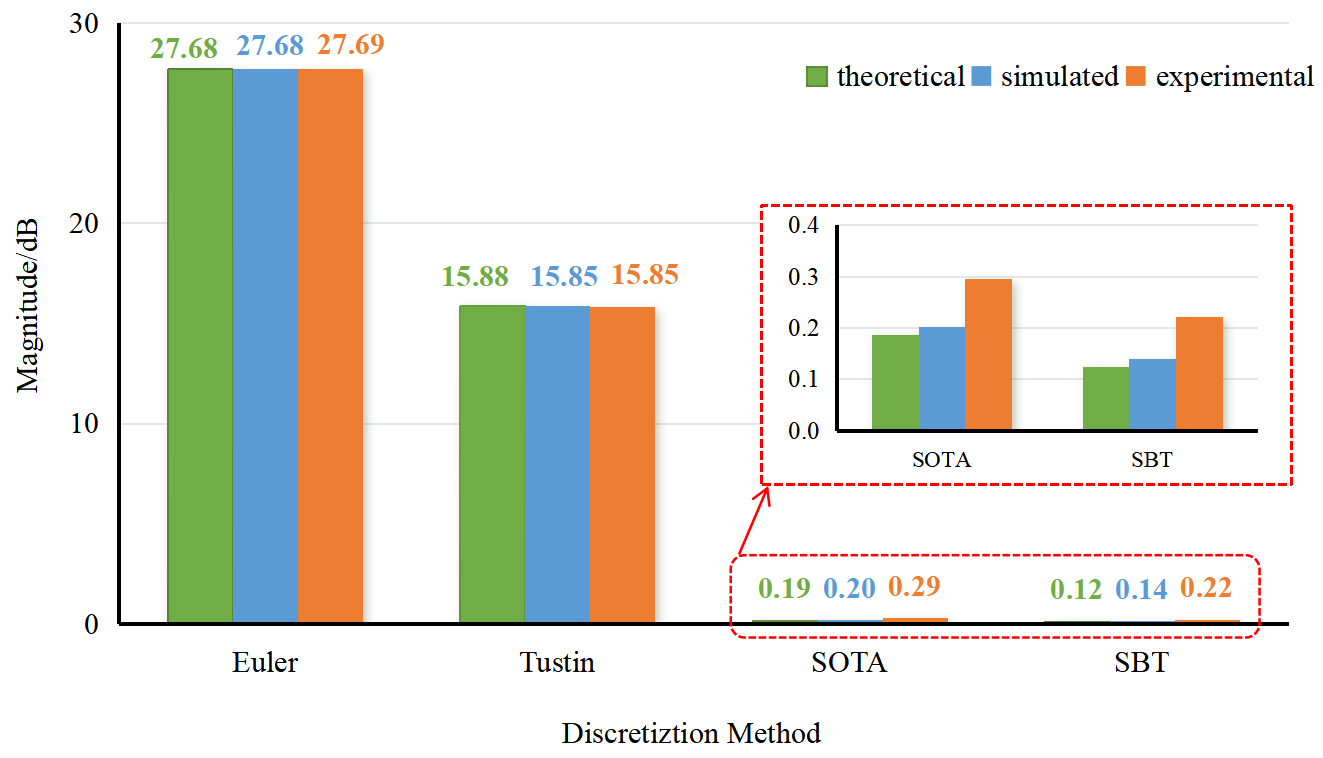}
  \caption{RMSE of magnitude comparison between theoretical analysis and experiments.}
  \label{fig_mag_rmse_comp}
\end{figure}

Fig.~\ref{fig_mag_rmse_comp} presents the root-mean-square error (RMSE) of the magnitude response, 
comparing theoretical predictions with simulation and experimental results across different discretization methods. 
The proposed SBT method achieves a substantial 25 \% reduction in RMSE compared to the SOTA approach. 
This improvement shows strong agreement with the theoretically predicted reduction of 33 \% and simulated reduction of 31 \%. 
The close correspondence among theory, simulation, and experiment conclusively validates the advancement of the proposed SBT method.


\subsection{Experiments on a Grid-tied Inverter}\label{sub2sec4}
\begin{figure}[htbp]  
  \centering
  \includegraphics[width=0.8\linewidth]{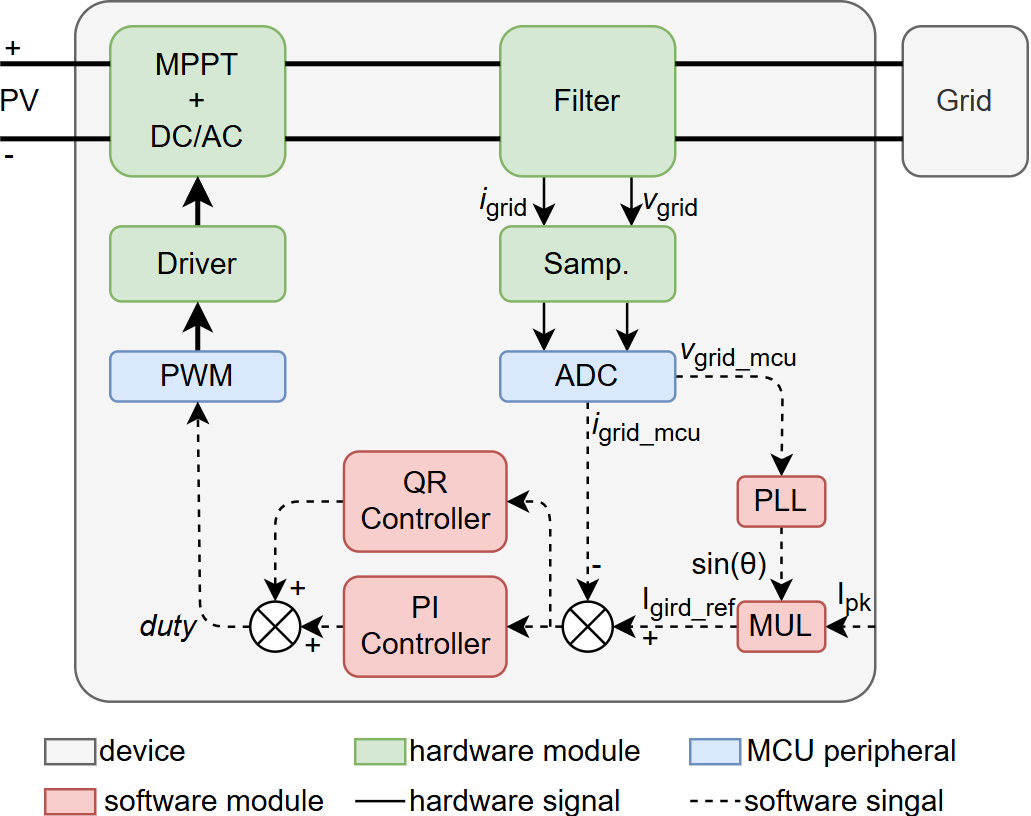}
  \caption{Block diagram of the gird-tied inverter test. Samp.: Sampling, PLL: Phase Lock Loop, MUL: Multiply, PWM: Pulse Width Modulation, ADC: A/D Conversion, DAC: D/A Conversion.}  
  \label{fig_inv_test_diagram}
\end{figure}

\begin{figure}[htbp]  
  \centering
  \includegraphics[width=1.0\linewidth]{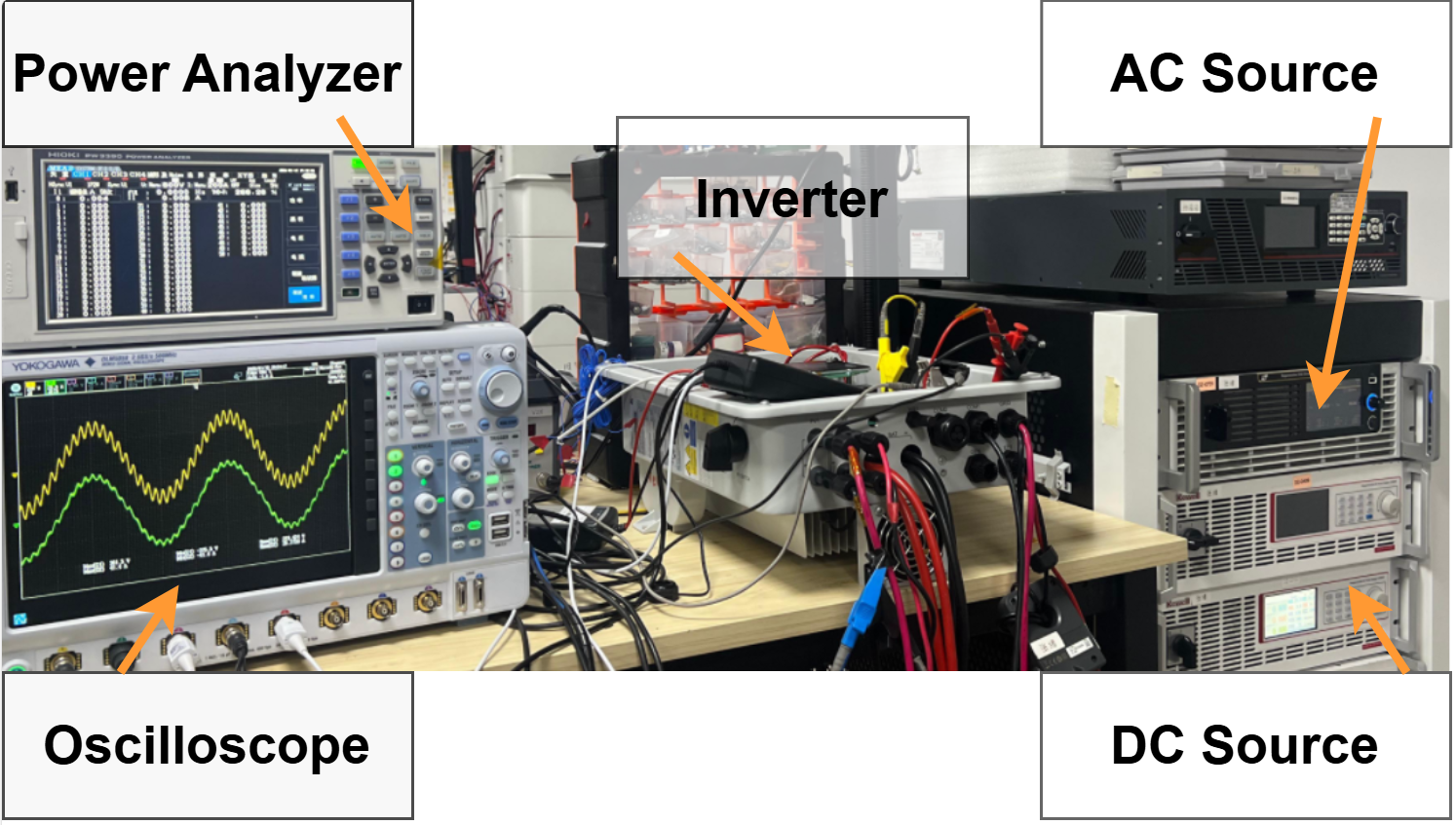}
  \caption{Physical setup of the gird-tied inverter test.}
  \label{fig_inv_test_setup}
\end{figure}

To further validate the advancement of the proposed SBT,
a grid-tied inverter is built as shown in Fig.~\ref{fig_inv_test_diagram}.
The control algorithms are implemented on a TMS320F28P65 microcontroller 
with a sampling frequency of 40 kHz (switching frequency is 20 kHz).
A DC source (Kewell C3000H) supplies the DC input.
The grid conditions, including harmonic distortions, 
are emulated using a programmable AC source (Chroma 61815), with the rated RMS setting to 220 V (50 Hz).
The total harmonic distortion (THD) of the AC voltage and current are measured using a high-precision power analyzer (HIOKI PW3390).
An LC filter, with parameters of 245 $\mu$H for inductance and 22 $\mu$F for capacitance, is incorporated in the inverter.
The waveforms are measured via an oscilloscope (YOKOGAWA DLM5058).
The physical setup of the grid-tied inverter test is shown in Fig.~\ref{fig_inv_test_setup}.

The controller used in this test is a PI+QR controller formulated as follows:
\begin{equation}
  G_{PIR}(s) = K_p(1+\frac{1}{\tau_i\cdot s}) + \frac{2K_r\cdot \omega_c\cdot s}{s^2 + 2\omega_c\cdot s + \omega_n^2}\label{eq_pir}
\end{equation} 

\begin{figure}[htbp]  
  \centering
  \includegraphics[width=0.8\linewidth]{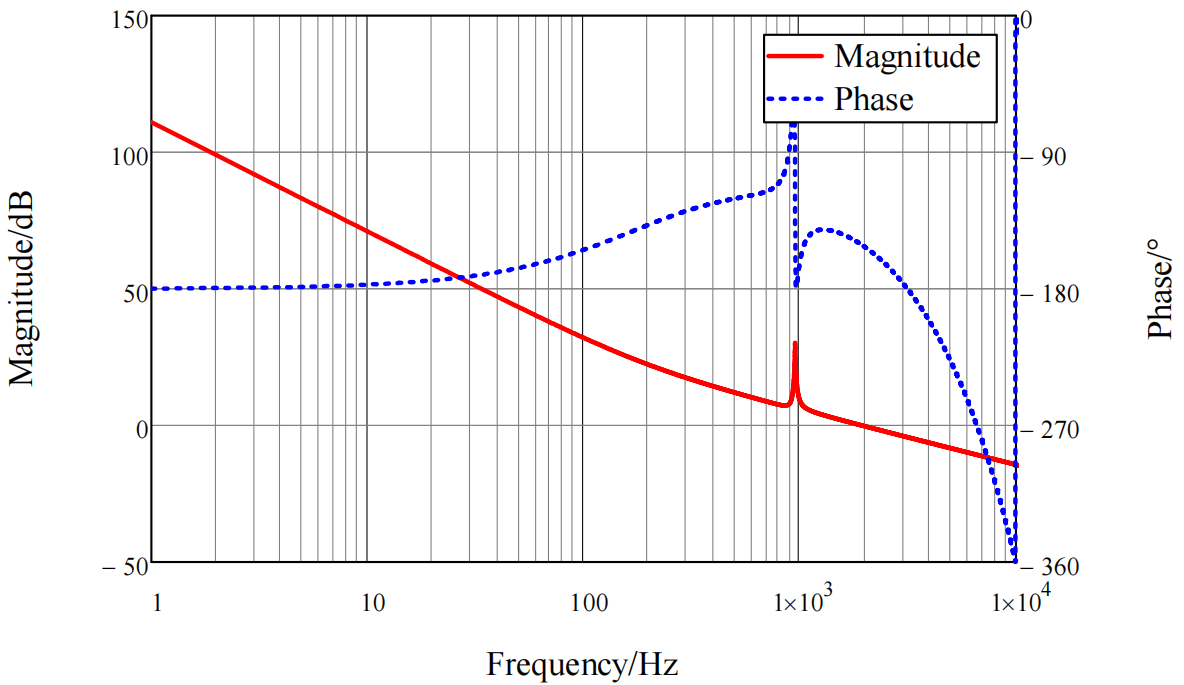}
  \caption{Bode diagram the grid-tied inverter.}
  \label{fig_inv_bode_plot}
\end{figure}

The controller parameters are optimized according to the method in \cite{PI09}, 
yielding $K_p=2.955$, $\tau_i=8.594\cdot10^{-4}$, $K_r=44.325$, $\omega_n=5969$ rad/s, and $\omega_c=17.907$ rad/s.
The Bode diagram of the optimized system, depicted in Fig.~\ref{fig_inv_bode_plot}, 
demonstrates a crossover frequency of 1.942 kHz, 
a phase margin of 28.8°, and a gain margin of 4.3 dB,
indicating adequate stability margins.

\begin{figure}[htbp]  
  \centering
  \includegraphics[width=1.0\linewidth]{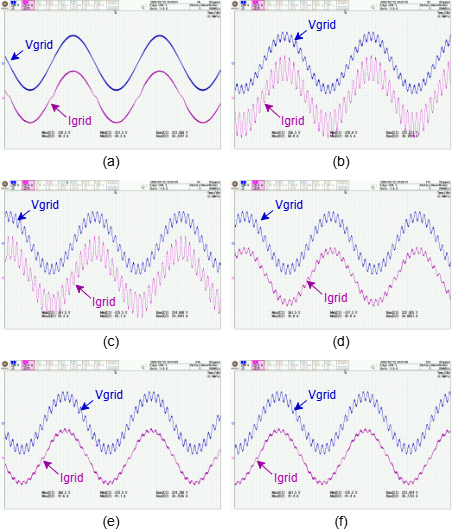}
  \caption{Experimental waveforms of the inverter test. (a) No harmonic injection; (b) PI controller alone; 
  (c) PI+QR via Euler; (d) PI+QR via Tustin; (e) PI+QR via SOTA; (f) PI+QR via SBT.
  CH1, grid voltage, 200 V/div, CH2, grid current, 25 A/div; time, 5 ms/div.}
  \label{fig_inv_test_waveform}
\end{figure}

The experimental waveforms captured at the resonant frequency are illustrated in Fig.~\ref{fig_inv_test_waveform}. 
Subfigures (a) and (b) depict the grid voltage and current under conditions without and with harmonic injection, respectively. 
Subfigure (b) corresponds to the case where only a PI controller is active. 
Subfigures (c) to (f) display the outcomes obtained with the PI+QR controller discretized via 
the Euler, Tustin, state-of-the-art (SOTA), and the proposed SBT methods, respectively. 
The amplitude of the harmonic injection is consistently maintained at 100 V (950 Hz) across all tests. 
The total harmonic distortion of the grid current (THDi) is quantified in Table~\ref{tab:thd_data}.
Analysis of the experimental waveforms and data indicates that the Euler method leads to severe resonance damping, 
resulting in a THDi of 37.92 \%—a level comparable to that achieved using the PI controller alone (38.73 \%).
The Tustin method yields a notable improvement, reducing THDi to 10.78 \%. 
The SOTA method further enhances performance, achieving an average THDi of 5.54 \%. 
In contrast, the proposed SBT method attains the best performance, with the lowest THDi of 5.44 \%. 
These results collectively verify the effectiveness of the proposed SBT method in the grid-tied inverter application.

\begin{table}[htbp] 
\caption{Experimental data of Total Harmonic Distortion\label{tab:thd_data}}
\centering
\begin{tabular}{|c|c|c|}
\hline
\textbf{Controller} & \textbf{Method} & \textbf{Experimental THDi}\\
\hline
PI     & Euler  & $38.73\pm 0.06$ \% \\
PI+QR  & Euler  & $37.92\pm 0.06$ \% \\
PI+QR  & Tustin & $10.78 \pm 0.05$ \% \\
PI+QR  & SOTA   & $5.54\pm 0.04$ \% \\
PI+QR  & SBT    & $5.44\pm 0.04$ \% \\
\hline
\end{tabular}
\end{table}

\section{Conclusion}\label{sec5}
Discretization is an essential step in the implementation of digital control systems, 
and the selection of an appropriate discretization method is critical to achieving high performance and accuracy.
This article introduces a novel bilinear transformation, 
termed the $\alpha\beta$-approximation or Scalable Bilinear Transformation (SBT), 
and demonstrates that the SBT serves as a unified framework for numerical integration.
The proposed method is also demonstrated to be highly effective for discretizing resonant controllers. 
The principal conclusions of this work are summarized as follows:
\begin{enumerate}
  \item The proposed SBT method utilizes a novel Scalable Hexagonal Approximation (SHA) 
  that approximates the error function's enclosed area with a scalable hexagonal shape. 
  Specifically, the two degrees of freedom are defined as the shape factor ($\alpha$) and the time factor ($\beta$).
  \item Two design methodologies for the parameters $\alpha$ and $\beta$ in the SBT 
  when discretizing resonant controllers are proposed: a straightforward approach and an optimal approach.
  \item The proposed SBT method delivers the optimal performance in discretizing resonant controllers, 
  as evidenced by a comprehensive comparison with the Euler, Tustin, and SOTA methods. 
  The performance variation is attributed to the equivalent pole position accuracy achieved through pole mapping. 
  Crucially, the SBT method accomplishes this enhanced performance at a computational cost equivalent to that of the simpler Euler and Tustin methods. 
\end{enumerate}

\section*{Acknowledgments}
This work was supported by the National Key Research and Development Program of China under Grant 2023YFB2604600.

\newpage
\section{Biography Section}

\begin{IEEEbiography}[{\includegraphics[width=1in,height=1.25in,clip,keepaspectratio]{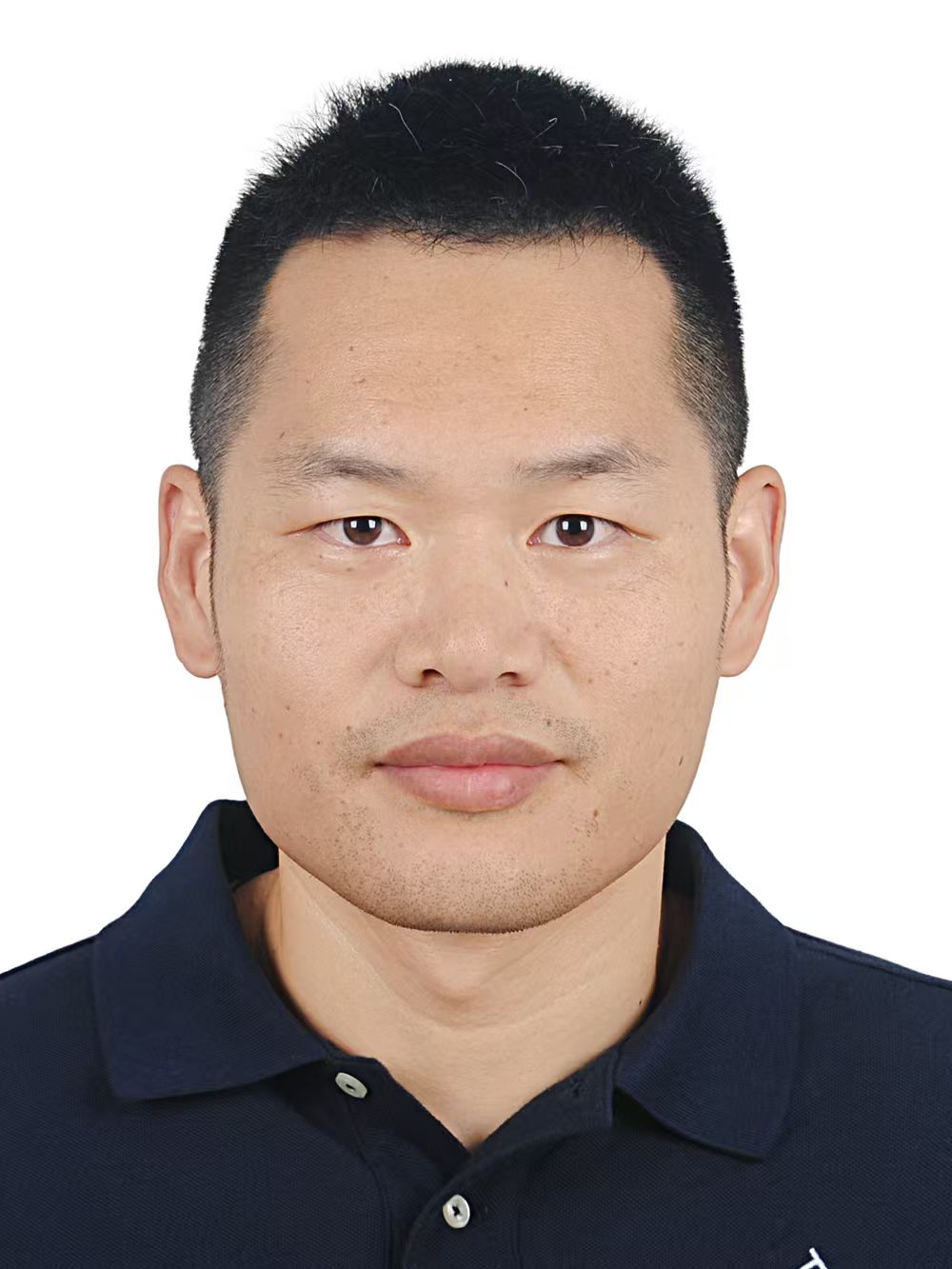}}]{Shen Chen}
(Student Member, IEEE) received the
B.S. degree in electrical engineering from Fuzhou university, Fuzhou, China, 
and the M.S. degree in power electronics and electric drive from Zhejiang university, Hangzhou, China, in 2009, and 2012, respectively.
He is currently working toward the Ph.D. degree with Xi'an Jiaotong University, Xi'an, China,
and a software leader with SolaX Power Network Technology (Zhejiang) Co., Ltd., Hangzhou, China.

His research interests include modeling and control of converters, system design of microgrids, embedded software engineering, self-adaptive control, and machine learning.
\end{IEEEbiography}

\begin{IEEEbiography}[{\includegraphics[width=1in,height=1.25in,clip,keepaspectratio]{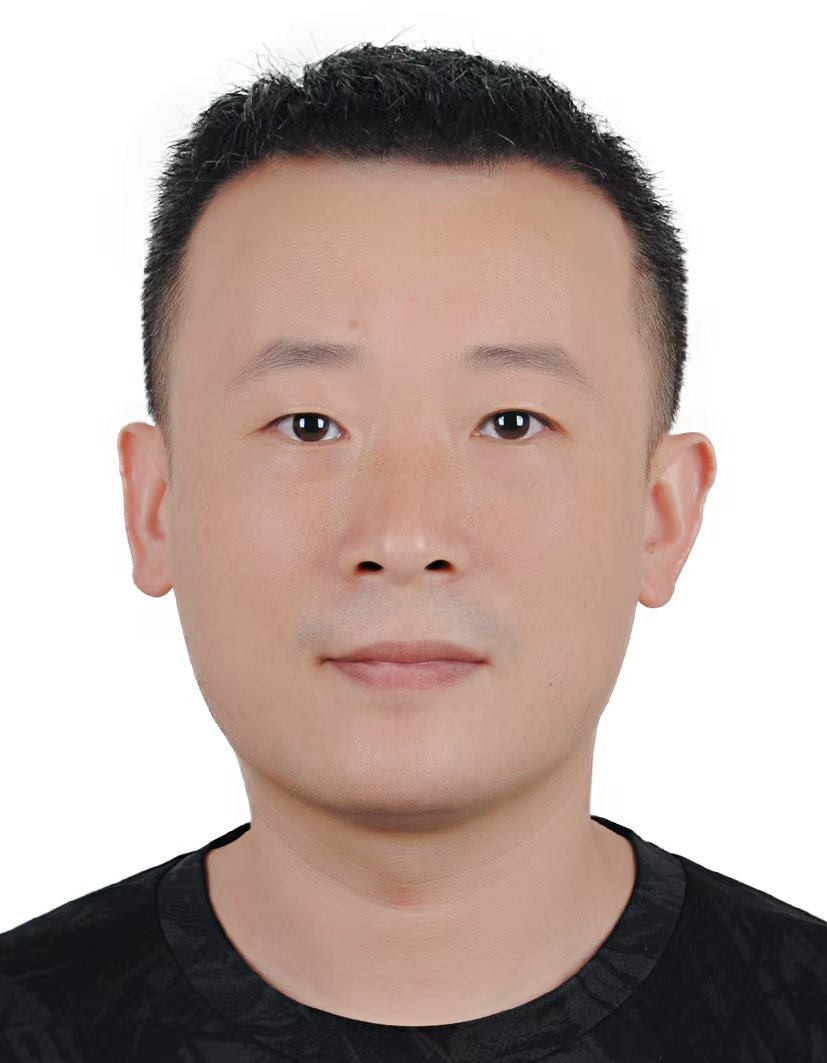}}]{Chaohou Liu}
received the B.S. degree in electrical engineering and automation from Qingdao Agricultural University, Qingdao, China, 
and the M.S. degree in power electronics and electric drive from Shanghai Maritime University, Shanghai, China, in 2008, and 2010, respectively.
He is currently a software leader with SolaX Power Network Technology (Zhejiang) Co., Ltd., Hangzhou, China.

His research interests include micogrid system, energy storage solutions and digital control in power electronics.
\end{IEEEbiography}

\begin{IEEEbiography}[{\includegraphics[width=1in,height=1.25in,clip,keepaspectratio]{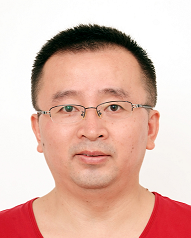}}]{Wei Yao} 
received the M.S degree in Control Theory and Control Engineering from Zhejiang University of Technology 
and the Ph.D. degree in Electric Engineering from Zhejiang University, in 2002 and 2015, respectively. 
He is currently an associate professor with College of Electric Engineering, 
Zhejiang University of Water Resources and Electric Power, Hangzhou, China.

His research interests include motor control, and power electronics, etc.
\end{IEEEbiography}

\begin{IEEEbiography}[{\includegraphics[width=1in,height=1.25in,clip,keepaspectratio]{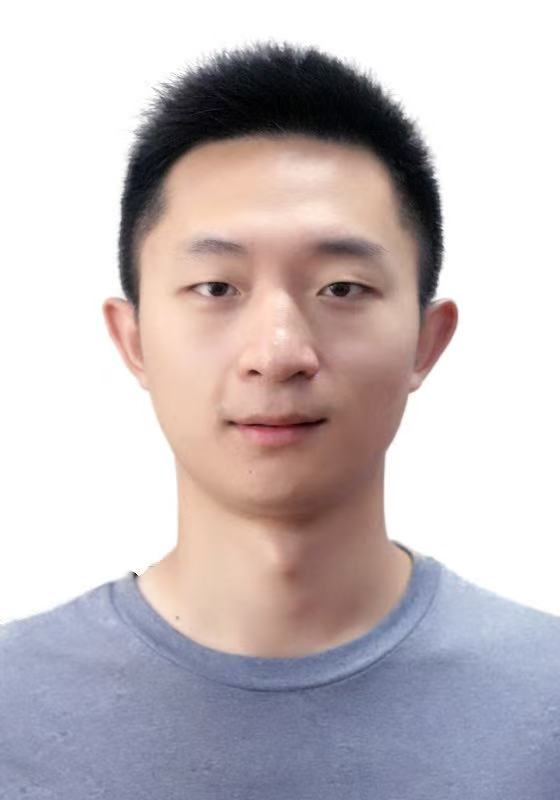}}]{Jisong Wang}
received the B.S. degree in electrical engineering from Yanshan University, Qinhuangdao, China, 
and  the M.S. degree from Naval University of Engineering, Wuhan, China, in 2016, and 2019, respectively.
He is currently a software manager with SolaX Power Network Technology (Zhejiang) Co., Ltd., Hangzhou, China.

His research interests include modeling, and control of inverters.
\end{IEEEbiography}

\begin{IEEEbiography}[{\includegraphics[width=1in,height=1.25in,clip,keepaspectratio]{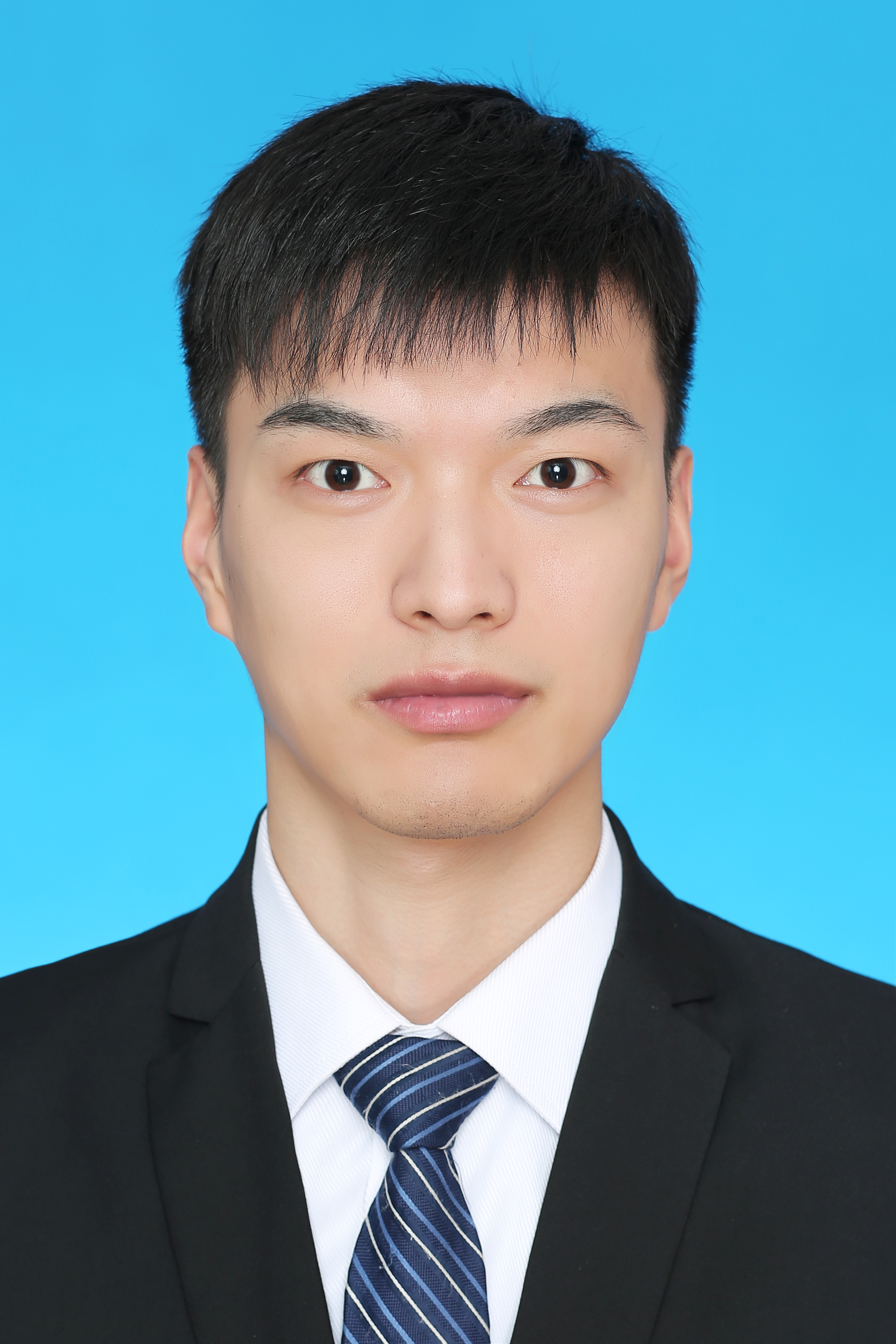}}]{Shuaipo Guo}
received the B.S. degree in electrical engineering and automation from Jiangsu University, Zhenjiang, China, 
and the M.S. degree in electrical engineering from the same university, in 2018, and 2021, respectively. 
He is currently a senior software engineer with SolaX Power Network Technology (Zhejiang) Co., Ltd., Hangzhou, China.

His research interests include modeling, and control of inverters.
\end{IEEEbiography}

\begin{IEEEbiography}[{\includegraphics[width=1in,height=1.25in,clip,keepaspectratio]{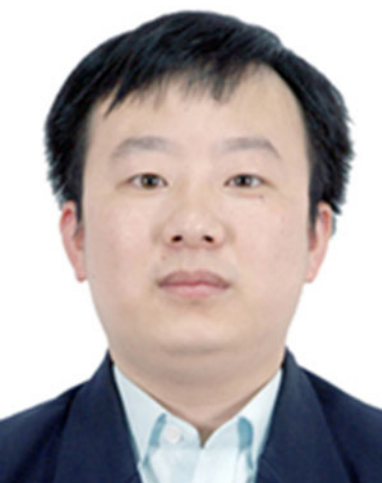}}]{Zeng Liu}
 (Member, IEEE) received the B.S. degree from Hunan University, Changsha, China, and the
M.S. and Ph.D. degrees from Xi'an Jiaotong University (XJTU), Xi'an, China, in 2006, 2009, and 2013,
respectively, all in electrical engineering.

He then joined XJTU as a Faculty Member in electrical engineering, where he is currently an Associate Professor. 
From 2015 to 2017, he was with the Center for Power Electronics Systems, 
Virginia Polytechnic Institute and State University, Blacksburg, VA, USA,
as a Visiting Scholar. 
His research interests include control of power systems with multiple converters for renewable energy and
energy storage applications, and small-signal stability of power electronics systems.

Dr. Liu was the recipient of two prize paper awards in IEEE TRANSACTIONS
ON POWER ELECTRONICS, and the CPSS Science and Technology Progress Award. 
He serves as an Associate Editor for the IEEE OPEN JOURNAL OF POWER ELECTRONICS and on the Editorial Board for the ENERGIES,
and served as Secretary-General for 2019 IEEE 10th International Symposium
on Power Electronics for Distributed Generation Systems, and 2020 the 4th International Conference on HVDC.

\end{IEEEbiography}

\begin{IEEEbiography}[{\includegraphics[width=1in,height=1.25in,clip,keepaspectratio]{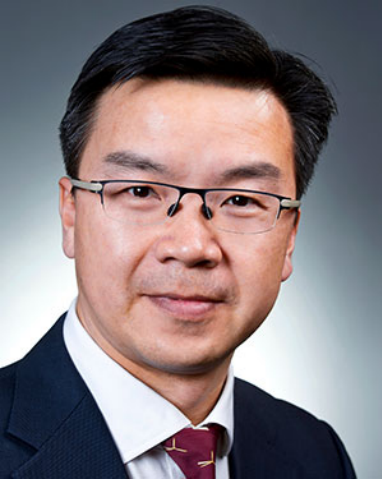}}]{Jinjun Liu}
(M’97–SM’10–Fellow’19) received the B.S. and Ph.D. degrees in electrical engineering from Xi'an Jiaotong University (XJTU), Xi'an, China, in 1992 and 1997, respectively.

He then joined the XJTU Electrical Engineering School as a faculty.
From late 1999 to early 2002, he was with the Center for Power Electronics Systems,
Virginia Polytechnic Institute and State University, Blacksburg, VA, USA, as a Visiting Scholar. 
In late 2002, he was promoted to a Full Professor and then the Head of the Power Electronics and Renewable Energy Center, XJTU, 
which now comprises over 30 faculty members and around 300 graduate students 
and carries one of the leading power electronics programs in China. 
From 2005 to early 2010, he served as an Associate Dean of Electrical Engineering School at XJTU, 
and from 2009 to early 2015, the Dean for Undergraduate Education of XJTU. 
He is currently a XJTU Distinguished Professor of Power Electronics. 
He coauthored 3 books (including one textbook), published over 500 technical papers in peer-reviewed journals and conference proceedings, 
and holds over 90 invention patents (China/US/EU). 
His research interests include modeling, control, and design methods and reliability evaluation and monitoring for power converters and 
electronified power systems, power quality control and utility applications of power electronics, 
and micro-grid techniques for sustainable energy and distributed generation.

Dr. Liu received for many times governmental awards at national level or provincial/ministerial level for scientific research/teaching achievements. 
He also received the 2006 Delta Scholar Award, the 2014 Chang Jiang Scholar Award, 
the 2014 Outstanding Sci-Tech Worker of the Nation Award, 
the 2016 State Council Special Subsidy Award, the IEEE Transactions on Power Electronics 2016 and 2021 Prize Paper Awards, 
the Nomination Award for the Grand Prize of 2020 Bao Steel Outstanding Teacher Award, 
the 2022 Fok Ying Tung Education and Teaching Award, and the 2025 IEEE PELS Harry A. Owen, Jr. Distinguished Service Award. 
He served as the IEEE Power Electronics Society Region 10 Liaison and then China Liaison for 10 years, 
an Associate Editor for the IEEE Transactions on Power Electronics since 2006, and 2015-2021 Vice President of IEEE PELS.
He was on the Board of China Electrotechnical Society 2012-2020 and was elected the Vice President in 2013 
and the Secretary General in 2018 of the CES Power Electronics Society. 
He was 2013-2021 Vice President for International Affairs, China Power Supply Society (CPSS), and since 2016, 
the inaugural Editor-in-Chief of CPSS Transactions on Power Electronics and Applications. 
He was elected the President of CPSS in Nov. 2021. 
Since 2013, he has been serving as the Vice Chair of the Chinese National Steering Committee for College Electric Power Engineering Education Programs.
\end{IEEEbiography}


\begin{thebibliography}{99}
\bibliographystyle{IEEEtran}

\bibitem{Direct-01}
S. Chen, Q. Zhao, Y. Ye, and B. Qu.
``Using IIR Filter in Fractional Order Phase Lead Compensation PIMR-RC for Grid-Tied Inverters,'' 
\textit{IEEE Trans. Ind. Electron.}, vol. 70, no. 9, pp. 9399-9409, 2023.

\bibitem{Direct-02}
M. Hu, W. Hua, C. Cheng, Y. Wang, and C. Lu.
``Discrete-Time Frequency-Domain Disturbance Observer to Mitigate Harmonic Current in PMSM Drives and the Implementation With Reduced Delay,'' 
\textit{IEEE Trans. Power Electron.}, vol. 38, no. 8, pp. 9482-9493, 2023.

\bibitem{Direct-03}
X. Sun, Z. Xu, M. Pan, C. Sun, W. Pan, and G. Lei.
``Sensorless Control Strategy for SRM Based on Flux Linkage in Medium to High-Speed Range,'' 
\textit{IEEE Trans. Ind. Electron.}, vol. 72, no. 10, pp. 9922-9930, 2025.

\bibitem{Euler-01}
X. Yang, X. Xiong, Z. Zou, and Y. Lou.
``Semi-Implicit Euler Digital Implementation of Conditioned Super-Twisting Algorithm With Actuation Saturation,'' 
\textit{IEEE Trans. Ind. Electron.}, vol. 70, no. 8, pp. 8388-8397, 2023.

\bibitem{Euler-02}
X. Jiang, S. Yang, X. Weng, Z. Wei, J. Liu, and Z. Zhang.
``Research on Open Circuit Fault Diagnosis Strategy for DFPMM Drive System Based on Grey Prediction Theory With the Introduction of Euler Algorithm,'' 
\textit{IEEE Trans. Ind. Electron.}, vol. 72, no. 3, pp. 2169-2179, 2025.

\bibitem{Heun-01}
D. Prajapati, A. Dekka, D. Ronanki, and J. Rodriguez.
``Low-Complexity Heun's Method-Based FCS-MPC With Reduced Common-Mode Voltage for a Five-Level Inverter,'' 
\textit{IEEE Trans. Power Electron.}, vol. 39, no. 3, pp. 3329-3338, 2024.

\bibitem{Tustin-01}
M. Wang, K. Kang, C. Zhang, and L. Li.
``Precise Position Control in Air-Bearing PMLSM System Using an Improved Anticipatory Fractional-Order Iterative Learning Control,'' 
\textit{IEEE Trans. Ind. Electron.}, vol. 71, no. 6, pp. 6073-6083, 2024.

\bibitem{Tustin-02}
Z. Sun, G. Ren, S. Xu, G. Ma, and J. Jatskevich.
``Six-Step Operation With Multistep Predictive Control Using the Trapezoidal Method for Traction PMSM Drives,'' 
\textit{IEEE Trans. Power Electron.}, vol. 39, no. 3, pp. 3486-3497, 2024.

\bibitem{SOTE-01}
G. Bi, Y. Chen, K. Wang, S. Cao, and T. Yang.
``Sensorless Control Design and Stability Analysis of High-Speed Dual Three-Phase Permanent Motor Drive for Future Aircraft Applications,'' 
\textit{IEEE Trans. Power Electron.}, vol. 40, no. 11, pp. 16169-16183, 2025.

\bibitem{HOTE-01}
Z. Fu, Y. Zhang, and W. Li.
``Solving Future Nonlinear Equation System via ZNN and Novel General ILR3S Formula With Multitype Manipulator Applications,'' 
\textit{IEEE Trans. Ind. Electron.}, vol. 71, no. 10, pp. 12623-12633, 2024.

\bibitem{HOTE-RK-01}
Z. Cheng, L. Li, X. Liu, X. Bai, and J. Liu.
``Sensorless Control Based on Discrete Fractional-Order Terminal Sliding Mode Observer for High-Speed PMSM With LCL Filter,'' 
\textit{IEEE Trans. Power Electron.}, vol. 40, no. 1, pp. 1654-1668, 2025.

\bibitem{Exact-01}
H. Yang, A. Xu, Y. Zhang, and X. Chai.
``Error Analysis and Design of Sliding-Mode-Observer-Based Sensorless PMSM Drives Under a Low Sampling Ratio,'' 
\textit{IEEE Trans. Power Electron.}, vol. 39, no. 7, pp. 7783-7792, 2024.

\bibitem{Exact-02}
K. Bándy, and P. Stumpf.
``Model Predictive Control of LC Filter Equipped Surface-Mounted PMSM Drives Using Exact Discretization,'' 
\textit{IEEE Trans. Ind. Electron.}, vol. 72, no. 3, pp. 2369-2379, 2025.

\bibitem{Disc-Review}
S. Chen, et al., {\it{Optimized Design of the Generalized Bilinear Transformation for Discretizing Analog Systems,}} arXiv [Online]. Available: \url{https://doi.org/10.48550/arXiv.2511.03403}

\bibitem{RC-01}
Y. Zhong, N. Roscoe, D. Holliday, T. Lim, and S. Finney.
``High-Efficiency mosfet-Based MMC Design for LVDC Distribution Systems,'' 
\textit{IEEE Trans. Ind. Appl.}, vol. 54, no. 1, pp. 321-334, 2018.

\bibitem{RC-02}
M. Ketzer, and C. Jacobina.
``Virtual Flux Sensorless Control for Shunt Active Power Filters With Quasi-Resonant Compensators,'' 
\textit{IEEE Trans. Power Electron.}, vol. 31, no. 7, pp. 4818-4830, 2016.

\bibitem{RC-03}
X. Wang, P. Loh, and F. Blaabjerg.
``Stability Analysis and Controller Synthesis for Single-Loop Voltage-Controlled VSIs,'' 
\textit{IEEE Trans. Power Electron.}, vol. 32, no. 9, pp. 7394-7404, 2017.

\bibitem{RC-04}
A. Yepes, F. Freijedo, J. Doval-Gandoy, Ó. López, J. Malvar, and P. Fernandez-Comesaña.
``Effects of Discretization Methods on the Performance of Resonant Controllers,'' 
\textit{IEEE Trans. Power Electron.}, vol. 25, no. 7, pp. 1692-1712, 2010.

\bibitem{RC-05}
T. Chin, Y. Cho, J. Lim, and J. Kang, 
``Harmonic Rotating Magnetic Field Injection Methods for Torque Ripple Suppression in the Overmodulation Region of PMSMs,'' 
\textit{IEEE J. Emerging Sel. Top. Power Electron.}, pp. 1-14, 2025 (Early Access).

\bibitem{RC-06}
S. Fang, J. Meng, Y. Meng, Y. Wang, and D. Huang.
``Discrete-Time Active Disturbance Rejection Current Control of PM Motor at Low Speed Using Resonant Sliding Mode,'' 
\textit{IEEE Trans. Transp. Electrif.}, vol. 9, no. 3, pp. 4783-4794, 2023.

\bibitem{RC-07}
Y. Tao, Z. Zhu, Q. Xu, H. Li, and L. Zhu.
``Tracking Control of Nanopositioning Stages Using Parallel Resonant Controllers for High-Speed Nonraster Sequential Scanning,'' 
\textit{IEEE Trans. Autom. Sci. Eng.}, vol. 18, no. 3, pp. 1218-1228, 2021.

\bibitem{RC-08}
A. Yepes, F. Freijedo, O. Lopez, and J. Doval-Gandoy.
``High Performance Digital Resonant Controllers Implemented with Two Integrators,'' 
\textit{IEEE Trans. Power Electron.}, vol. 26, no. 2, pp. 563-576, 2011.

\bibitem{RC-09}
M. Hu, W. Hua, W. Huang, and J. Meng.
``Digital Current Control of an Asymmetrical Dual Three-Phase Flux-Switching Permanent Magnet Machine,'' 
\textit{IEEE Trans. Ind. Electron.}, vol. 67, no. 6, pp. 4281-4291, 2020.

\bibitem{RC-10}
O. Husev, C. Roncero-Clemente, E. Makovenko, S. Pimentel, D. Vinnikov, and J. Martins.
``Optimization and Implementation of the Proportional-Resonant Controller for Grid-Connected Inverter With Significant Computation Delay,'' 
\textit{IEEE Trans. Ind. Electron.}, vol. 67, no. 2, pp. 1201-1211, 2020.

\bibitem{GBT-01-Sekara}
T. Šekara, and M. Stojic.
``Application of the $\alpha$-approximation for discretization of analog systems,''
\textit{Electrical Engineering}, vol. 18, no. 3, pp. 571-586, 2005.

\bibitem{GBT-02}
G. Zhang, X. Chen, and T. Chen.
``Digital redesign via the generalised bilinear transformation,''
\textit{International Journal of Control}, vol. 82, no. 4, pp. 741-754, 2009.

\bibitem{GBT-03}
T. Kim, J. Han, T. Oh, Y. Kim, S. Lee, and D. Dan Cho.
``A new accurate discretization method for high-frequency component mechatronics systems,''
\textit{Mechatronics}, vol. 62, pp. 102250, 2019.

\bibitem{PI09}
D. Homes, T. Lipo, B. McGrath, and W. Kong.
``Optimized Design of Stationary Frame Three Phase AC Current Regulators,'' 
\textit{IEEE Trans. Power Electron.}, vol. 24, no. 11, pp. 2417-2426, 2009.

\end{thebibliography}
\end{document}